\documentclass[mnsc,nonblindrev]{informs3} 
\usepackage{lineno}
\usepackage{amsmath}

\usepackage{amsmath}

\usepackage[T1]{fontenc}
\usepackage{threeparttable}
\usepackage{CJKutf8}
\usepackage{changes}
\usepackage{amssymb}
\usepackage{bbm}
\usepackage{amsmath}
\usepackage[colorlinks=true,allcolors=blue]{hyperref}%

\newtheorem{res}{Result}
\newtheorem{hyp}{Hypothesis}
\newcommand{\support}[1]{\noindent \textbf{Support.} \textit{#1}}

\usepackage{graphicx}

\usepackage{appendix}
\usepackage{natbib}
\usepackage{url}
\usepackage{setspace}
\usepackage{enumitem}
\newcommand{\yuan}[1]{\textcolor{brown}{#1 --YY}}

\newcommand{\reminder}[1]{\textcolor{red}{#1}}
\newcommand{\hide}[1]{} 

\newcommand{\mytablenote}{\textit{Note}: \linespread{1.25}\selectfont
The dependent variable (DV) for Columns (1) and (2) is the amount sent within the respective timeframe. It is coded as zero for those who do not send red packets.
The DV in Columns (3) and (4) is the dummy variable for sending red packets. 
The DV in Columns (5) and (6) is the amount conditioning on sending red packets. Marginal effects are reported.
Standard errors clustered at the group  and user level are in parentheses. 
*: $p<0.1$, **: $p<0.05$, ***: $p<0.01$. 
}

\OneAndAHalfSpacedXI 

\usepackage{natbib}
 \bibpunct[, ]{(}{)}{,}{a}{}{,}%
 \def\bibfont{\small}%
\MANUSCRIPTNO{MS-DEC-19-02859}

\begin{document}



\RUNTITLE{Gift Contagion in Online Groups}

\TITLE{
Gift Contagion in Online Groups: Evidence From Virtual Red Packets}
\ARTICLEAUTHORS{%
\AUTHOR{Yuan Yuan}
\AFF{Mitchell E. Daniels, Jr. School of Business, Purdue University} 
\AUTHOR{Tracy Xiao Liu}
\AFF{Department of Economics, School of Economics and Management, Tsinghua University}
\AUTHOR{Chenhao Tan}
\AFF{Department of Computer Science \& Harris School of Public Policy, University of Chicago} 
\AUTHOR{Qian Chen}
\AFF{Tencent Inc.} 
\AUTHOR{Alex `Sandy' Pentland}
\AFF{Media Lab, Massachusetts Institute of Technology} 
\AUTHOR{Jie Tang}
\AFF{Department of Computer Science, Tsinghua University}

}

\ABSTRACT{
Gifts are important instruments for forming bonds in interpersonal relationships. Our study analyzes the phenomenon of gift contagion in online groups. Gift contagion encourages social bonds by prompting further gifts; it may also promote group interaction and solidarity. Using data on 36 million online red packet gifts on a large social site in East Asia, we leverage a natural experimental design to identify the social contagion of gift giving in online groups. Our natural experiment is enabled by the randomization of the gift amount allocation algorithm on the platform, which addresses the common challenge of causal identifications in observational data. Our study provides evidence of gift contagion: on average, receiving one additional dollar causes a recipient to send 18 cents back to the group within the subsequent 24 hours. Decomposing this effect, we find that it is mainly driven by the extensive margin -- more recipients are triggered to send red packets. Moreover, we find that this effect is stronger for “luckiest draw” recipients, suggesting the presence of a group norm regarding the next red packet sender. Finally, we investigate the moderating effects of group- and individual-level social network characteristics on gift contagion as well as the causal impact of receiving gifts on group network structure. Our study has implications for promoting group dynamics and designing marketing strategies for product adoption. 
}%



\maketitle

%


\section{Introduction}

Individuals belong to many different social groups: kinship groups, friend groups, work groups or organizations, and interest groups. 
The collective identities developed in these groups deeply shape the behavior of their members \citep{tajfel1979integrative, cialdini2004social,christakis2007spread,chen2009group,aral2012identifying,bond201261}. Nowadays, social groups are facilitated through digital platforms, especially social network platforms such as Facebook, Line, WeChat, and WhatsApp. These platforms support social groups for a variety of purposes, including relationship maintenance, opinion and information exchange, and event planning \citep{veinott1999video,backstrom2006group,park2009being,bloom2015does,liu2015enterprise}. 
In particular, during the COVID-19 pandemic, online work group chats have substituted for conventional in-person meetings \citep{brynjolfsson2020covid}; indeed,
it was reported that 42 percent of the U.S. labor force worked from home full-time as of June 2020.\footnote{\url{https://news.stanford.edu/2020/06/29/snapshot-new-working-home-economy/}} 
In China, WeChat groups are widely used for instant work-related communication \citep{liu2015enterprise,qiu2016lifecycle}. 
Although online work groups offer the convenience of long-distance communication and coordination, they may face challenges related to team building and group solidarity \citep{holton2001building}.






One way to promote group bonding is through the use of group gifts, which are the gifts sent by a group member without specifying recipients. 
Examples of group gifts include the food items or souvenirs bought by a member to her work group after traveling abroad as well as the small gifts being exchanged at a Christmas or holiday party (the ``white elephant'' gift exchange). 
While prior literature focuses on one-to-one gifts and their role in creating interpersonal social bonds \citep{mauss2002gift}, few studies have investigated group gifts and their role in promoting in-group interactions and solidarity.

In this study, we are interested in studying the outbreak of sending group gifts in online social groups, indicating the presence of gift contagion (the social contagion of gift giving).
Social contagion is defined as ``{an event in which a recipient's behavior has changed to become `more like' that of the actor}'' \citep{wheeler1966toward}. \cite{aral2009distinguishing} has pointed to the importance of identifying causal effects in the process of social contagion. 
In the context of group gifts, gift contagion implies that people who receive larger amounts of gifts feel promoted to increase their own subsequent contributions.
If gift contagion exists in groups, the actual impact of a given gift would be amplified, leading to 
stronger social bonds and feelings of  group solidarity \citep{markovsky1994new}.
To quantify the effect of gift contagion, our study employs a large-scale dataset of 3.4 million users on a large social network platform in east Asia. For anonymous concern, we call the platform we study as ABC thereafter. 
On the platform, users send online red packets to each other as a type of digital monetary gift. 
The red packets, especially group red packets, swiftly became extremely popular after being released. For example,  
more than 700 million people engaged in sending or receiving red packets during one week in 2019. 



Methodologically, the causal identification of social contagion in observational data is notoriously challenging \citep{aral2009distinguishing,shalizi2011homophily,yuan2021causal}. In particular,   the following two confounding factors may hinder valid causal identification of gift contagion. 
The first confounding is the ``temporal clustering.'' Specifically, group members  may send gifts within a short time period independently   to celebrate a festival or an event \citep{aral2009distinguishing}. 
The second is homophily, the phenomenon whereby individuals tend to befriend similar others \citep{mcpherson2001birds}.
For example, wealthy people tend to cluster in a group and send larger amounts of gifts to each other. 

We leverage a natural experiment to overcome the above challenges. 
Our natural experiment is enabled by a random gift amount allocation algorithm for group red packets.
The algorithm splits a red packet into several shares and randomly determines the amount of each share. 
We utilize this random assignment of received cash amounts to identify the impact of the amount received on a participant's subsequent gifting behavior.
We first examine the presence of gift contagion on the platform. 
On average, receiving one additional dollar causes a recipient to send 18 cents back to the group within the subsequent 24 hours. 
Moreover, we find that this overall effect is mainly driven by the extensive margin, i.e., receiving red packets significantly promotes the likelihood of giving. 
Second, we investigate heterogeneity in the effect size of gift contagion across different time periods (festival versus non-festival periods), and across different types of groups (e.g., relative versus classmate groups).
Third, our analysis suggests that  a social norm exists in that the luckiest draw recipient should send the very first subsequent red packet.
Finally, we find that both individual-level network position and group-level network structure  affect the strength of gift contagion, respectively.

The rest of the paper is organized as follows.
Section~\ref{sec:data} describes the institutional background and our data. 
Section~\ref{sec:strategy} discusses our estimation strategy. 
Section~\ref{sec:hyp} introduces our hypotheses. 
Section~\ref{sec:result} presents the results. 
Section~\ref{sec:dis} concludes. 

\section{Data}\label{sec:data}

\subsection{Background}
\begin{figure}
    \centering
    \includegraphics[width=0.8\linewidth]{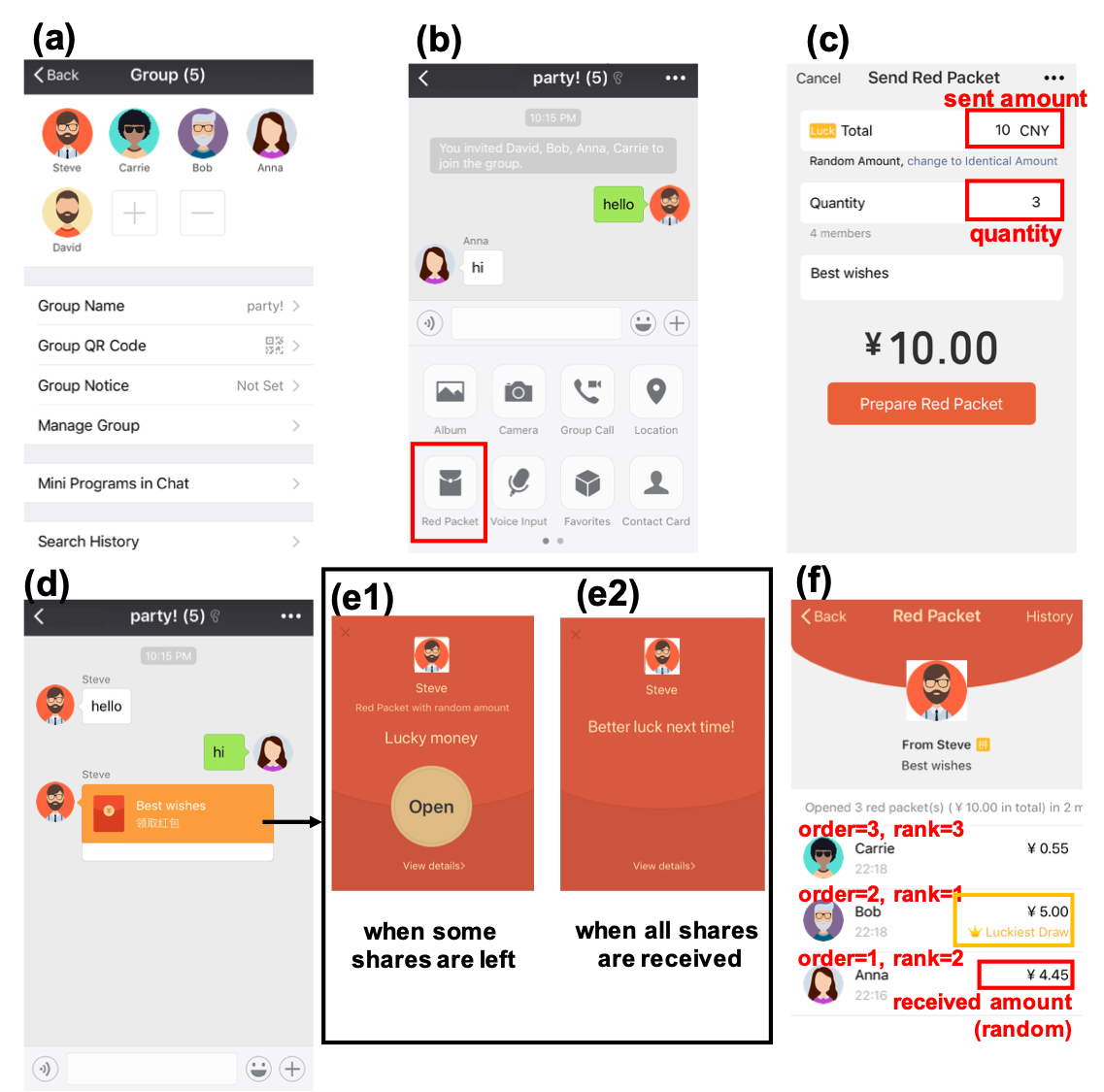}
    \caption{Illustration of group red packets with random amounts }
    \label{fig:illustration}
\end{figure}

In our study, we focus on the custom of sending monetary gifts to family members, friends, and other acquaintances in what is known as ``red packets'' (also known as ``red envelopes,'' or ``lucky money''). 
Red packets are typically sent to others as a way of commemorating festivals or important events. 
They also function as a means of tightening the social network in East and Southeast Asian cultures, known as the \textit{renqing} and \textit{guanxi} system \citep{luo2008changing,wang2008significance}. We summarize the history of red packets in \textit{Appendix~A}.

ABC platform enables users to designate private contacts (we use the term ``friends'' throughout the paper) and to create group chats.  Groups are created for a wide variety of purposes, ranging from family members to coworkers  and friends. The number of group members ranges between 3 and 500. 
ABC introduced its online red packet feature in 2014, allowing users to send monetary gifts to either a friend or a group. Its red packets were popularized during the Lunar New Year of 2015 --- it was reported that 55\% of the Chinese population sent and received red packets on that single day. 
Because of the popularity of red packets, ABC also benefits from a rapid growth in its mobile payment market share and becomes the second largest mobile payment platform in the country. 

Noting there are two types of group red packets, our study focuses on the most popular one -- the random-amount, group-designated red packet, as depicted in Figure~\ref{fig:illustration}.\footnote{The other type is that senders can also choose to split red packets equally.} 
In this example, 
Panel (a) provides the basic information about the group, the name of which is “Party!”, with five group members in total.
Panel (b) presents the user interface of this group, from which a group member can click the ``Red Packet'' button to send a red packet. Panel (c) shows that a group member (Steve in this example) can choose both the \textit{total amount} of the red packet that he would like to send (``Total''$=$10 CNY) and the \textit{number of recipients} (``Quantity''$=$3). 
Panel (d) is the interface for the red packet notification, from which a user can click the orange button to choose to receive the packet. Panel (e1) pops up when some shares of the red packet remain. In this example, only the first three users who click the ``Open'' button can receive a share of this red packet. Panel (e2) pops up when all shares of the red packet have been received by group members. Finally, Panel (f) shows the recipient list, which can be viewed by clicking ``View details'' in Panel (e). All of the group members, including senders, recipients, and non-recipients, can view the recipient list and see the amount obtained by each recipient. We define the \textit{order of receiving time} as group members' respective places in the order at the time when they receive the red packet. In the above setting, the amount that each user receives is randomly assigned by the platform, and is a function of the total amount of the red packet, the number of recipients, and the order of the receiving time.
Moreover, the platform designates which user receives the largest amount of a red packet with a ``Luckiest Draw'' icon and the corresponding yellow text. All group members can observe who is the luckiest draw recipient.

\subsection{Data collection}
In collaboration with the company, we collect a dataset consisting of randomly-sampled groups with red packet activity from October 1, 2015 to February 29, 2016. To protect user privacy, users' identities were anonymized before we accessed the data. To avoid data sparsity, we restrict our analysis to  groups in which the number of red packets sent is at least three times the number of group members. 
We also filter out groups that might be used for online gambling based on the following criteria: (1) a number of red packets greater than 50 times the number of group members; (2) a name that suggests a gambling focus (containing words such as  \begin{CJK*}{UTF8}{gbsn}元/块(Chinese yuan), 发(send), 红(red), 包(packet), 最佳(luckiest), 抢(grad), 赌(gamble), 钱(money), 福利(welfare), and 接龙(chain)\end{CJK*}  or Arabic numerals (which indicate the default packet amount set for gambling); or (3) no designated group name, which could also be temporarily created for gambling.\footnote{We show that groups identified as gambling groups appear to exhibit greater levels of gift contagion (\textit{Appendix~D.1}).} 
In total, this selection process results in 174,131 groups with 3,466,928 group members (3,450,540 unique users). 

In our main analyses, we include: (1) the characteristics of 174,131 groups, including the number of group members, the total number of red packets, and the total cash value of the red packets; (2) 3,450,540 unique users in these groups, along with their characteristics, such as the number of in-group friends are also retrieved; and (3) the attributes of each red packet, including the cash amount, the corresponding recipients and the opening time. In total, our sample comprises 36,608,864 red packets. Furthermore, we focus on recipients of ``spontaneous red packets,'' which indicate that no group red packet is sent in the 24 hours prior to this type of red packet. We conduct robustness checks by varying the time window and find that our main results remain robust (\textit{Appendix~D.1}).
In total, we identify 1,549,720 spontaneous red packets sent to 7,266,446 recipients.\footnote{We exclude observations in which the sender clicks the red packet and receives a share of her own red packet.} 
Each observation refers to a user's received red packet and we have 7,266,446 observations in total.  \textit{Appendix~B} presents a detailed description of the data.

\section{Estimation strategy}
\label{sec:strategy}

\subsection{Random assignment algorithm for group red packets}

Here we illustrate the random assignment algorithm for red packet amounts. 
First, the sender determines the total amount of the red packet ($a>0$) and the number of recipients to receive a portion of the red packet ($n\geq 1$).\footnote{In practice, the amount received is rounded to the nearest cent, and is set at least 0.01 CNY.}  Then group members choose to open the red packet on a first-come, first-served basis. They do not know the values of $a$ and $n$ until they open the red packet. Let $o$ denote the order of receiving time ($o=1,2,...,n$). The amount received by the recipient with order $o$, denoted by $V_o$, is determined by the following algorithm:

\begin{enumerate}
    \item When $o=1$ and $o<n$: the amount obtained by the first recipient ($\text{order}=1$) follows a uniform distribution on (0, $\frac{2a}{n}$]. The expected amount is: $$\mathbbm{E}[V_1]= \frac{1}{2}\times (0+\frac{2a}{n}) = \frac{a}{n}.$$
    
    When $o=n=1$, the amount received is $a$ because the only recipient should take all the cash amount.
    
    \item When $1<o<n$: the amount received follows a uniform distribution on  (0, $\frac{2(a-V_1-...-V_{o-1})}{n-o+1}$]. 
    We have: 
    \begin{equation*}
    \begin{split}
    \mathbbm{E}[{V_o}]& = \mathbbm{E}\Big[\mathbbm{E}[V_o|V_1, ... ,V_{o-1}]\Big] \\
    & = \mathbbm{E} \bigg[\frac{1}{2} \times \Big(0 + \frac{2(a-V_1-...-V_{o-1})}{n-o+1} \Big)\bigg] \\
    & =   \frac{a-\mathbbm{E}[V_1]-...-\mathbbm{E}[V_{o-1}]}{n-o+1}.
    \end{split}
    \end{equation*}
    We show that $\mathbbm{E}[{V_o}]=\frac{a}{n}$ by induction:
    
    First, we have already shown that $\mathbbm{E}[V_1]=\frac{a}{n}$. 
    
    Second, assuming that we have $\mathbbm{E}[V_{o'}] = \frac{a}{n} $ for all $o'< o$, we have  $\mathbbm{E}[{V_o}] = \frac{a-(o-1)\frac{a}{n}}{n-o+1}  = \frac{a}{n}$.
   
    \item When $o=n$: $V_o=a-V_1 - ... - V_{o-1}$, indicating that the last recipient takes the surplus.  Then we have $\mathbbm{E}[{V_o}] =  a-\mathbbm{E}[V_1]-...-\mathbbm{E}[V_{o-1}]= \frac{a}{n}$. 
\end{enumerate}

Therefore, the expectation of the received amount is the same: $\frac{a}{n}$. However, the variance in the amounts is not always the same. For example, when $n>2$,
\begin{equation*}
\begin{split}
\text{Var}(V_1) & = \frac{1}{12} \times (\frac{2a}{n}-0)^2 = \frac{a^2}{3n^2}; \\
\text{Var}(V_2) & = \mathbbm{E}\Big[\text{Var}(V_2|V_1)\Big] + \text{Var}\Big(\mathbbm{E}[V_2|V_1]\Big) \\
& = 
\mathbbm{E}\Big[\frac{1}{12} \times \big(\frac{2(a-V_1)}{n-1}\big)^2\Big] + \text{Var}\Big(\frac{a-V_1}{n-1}\Big) \\
& = 
\mathbbm{E}\Big[\frac{(a-V_1)^2}{3(n-1)^2}\Big] + \frac{1}{(n-1)^2} \text{Var}(V_1) \\
& = \Big(- \frac{(a-\frac{2a}{n})^3}{9(n-1)^2} + \frac{a^3}{9(n-1)^2} \Big) \times \frac{n}{2a} + \frac{a^2}{3(n-1)^2n^2} \\ 
& = \frac{a^2}{3n^2}+\frac{4a^2}{9(n-1)^2n^2}> \text{Var}(V_1).
\end{split}
\end{equation*} 
In addition, we provide the complete proof for variance differences in \textit{Appendix~C}.



\begin{figure}[h!]
\centering
\includegraphics[height=2.6in]{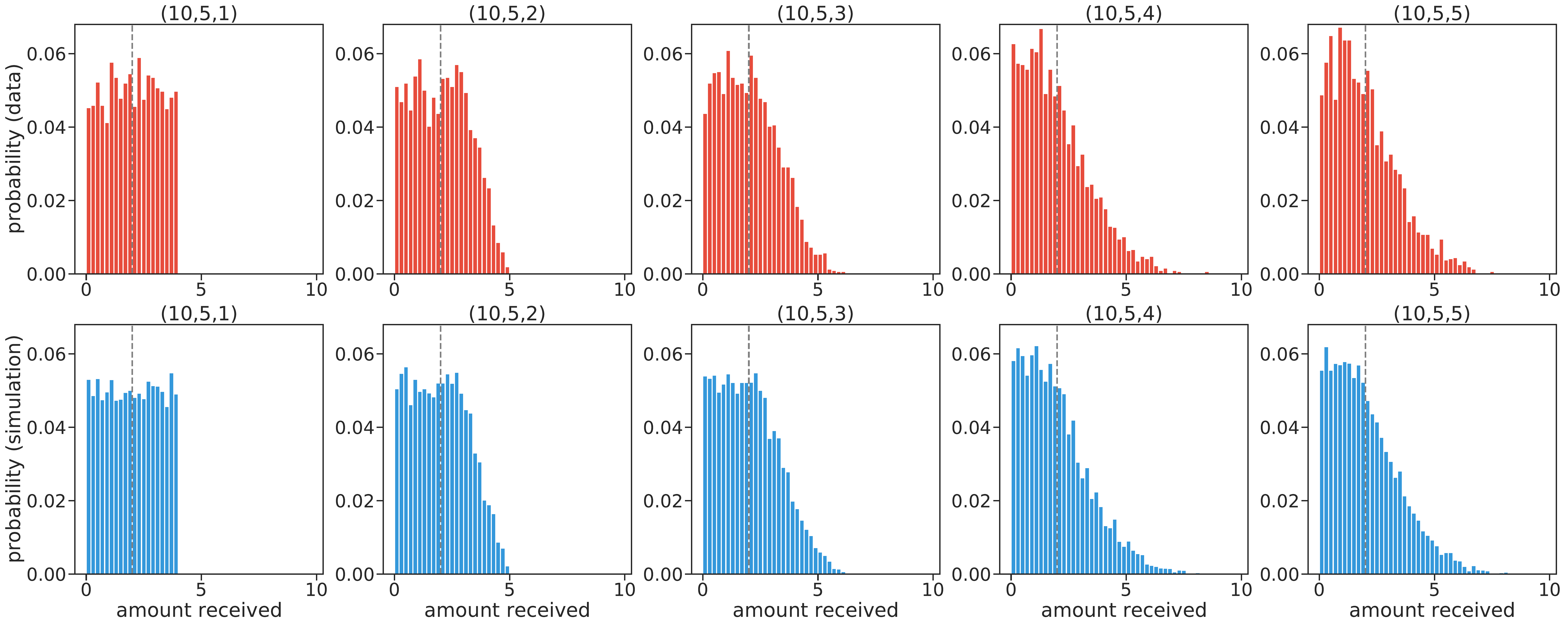}
\includegraphics[height=2.6in]{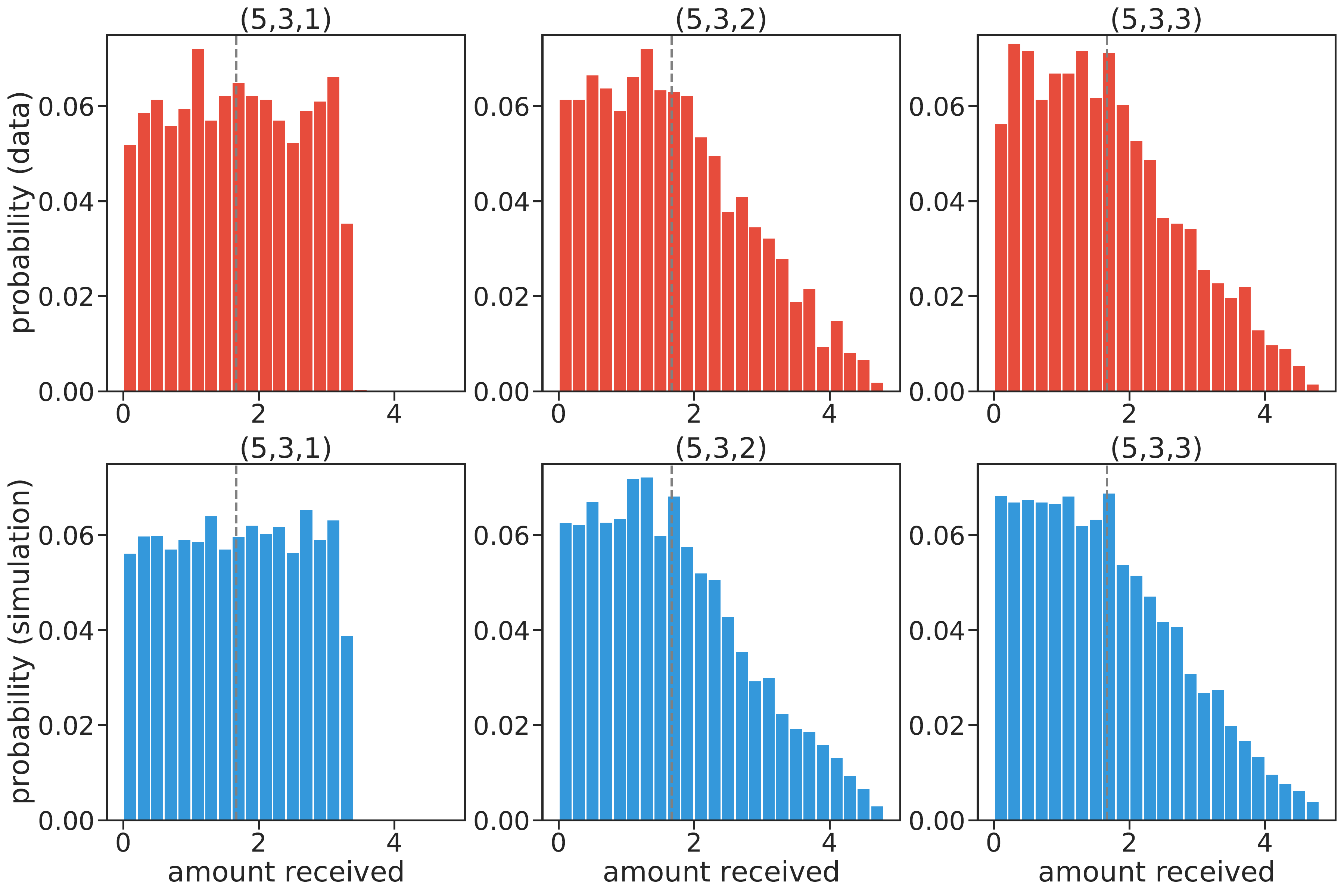}
\caption{\raggedright
\linespread{1.2} \selectfont  Distributions of received amounts in our dataset (red) and simulation (blue).
The top two rows are those with 10 CNY and 5  recipients; 	
The bottom two rows are those with 5 CNY and 3 recipients. A title with $(a,n,o)$ indicates that the total amount of the gift is $a$ CNY, the number of recipients is $n$, and the order of receiving time is $o$. 
\label{fig:order}
}
\label{fig:random1}
\end{figure}

To show that the random assignment algorithm implemented on the platform functions as described above, we compare the empirical distributions of received amounts from our data to the simulation results generated by the algorithm in Figure~\ref{fig:order}. In our first comparison example in the upper two rows, we see that the total amount is 10 CNY and the number of recipients is 5 (108,560 observations). In our second comparison example in the bottom two rows, we see that the total amount is 5 CNY and the number of recipients is 3 (38,523 observations). We do not find significant differences between these two distributions generated by the simulation and our empirical data ($p = 0.30$ and $0.36$ for the two cases, respectively, two-sided Kolmogorov-Smirnov tests).  
Additionally, consistent with the random assignment algorithm, the expectation of the amount received is solely determined by the total amount of the gift and the number of recipients ($\frac{10}{5}=2$ and $\frac{5}{3}$ for the two cases, respectively).
Examining the remaining data, we find that the results continue to hold. 

Furthermore, we verify the randomization procedure and provide the  results in \textit{Appendix~C}. The results suggest that, conditional on the total amount of the red packet, the number of recipients, and the order of receiving time, the amount received is not significantly correlated with individual characteristics or historical behaviors.  
Taken together, these analyses confirm that the amount that a recipient obtains is solely determined by the following three variables: (1) the total amount of the red packet; (2) the number of recipients; and (3) the order of receiving time. This verification enables us to use the following empirical strategy to quantify the causal impact of gift contagion. 

\subsection{Empirical strategy} 

\label{sec:empirical}
We next discuss our empirical strategy, which is used to quantify the impact of the amount received on the recipient’s subsequent gifting behavior. We regard the random assignment of received amounts as a stratified randomized experiment \citep{kernan1999stratified,imai2008misunderstandings,imbens2015causal,athey2017econometrics}, where a stratum is uniquely determined by the total amount of the red packet, the number of recipients, and the order of receiving time.  We apply the empirical strategy of stratified randomized experiments proposed by \cite{imbens2015causal} and conduct the following regression analyses: 

\begin{equation}
Y_{gir} = \beta T_{gir} + \sum_s \gamma_s B_s({A_{r}, N_{r}, O_{ir}})+ \epsilon_{gir}.
\label{eq:all}
\end{equation}

In Eq.~(\ref{eq:all}), $g$ denotes a group, and $i$ denotes a unique user who receives a share of a red packet $r$. $\epsilon_{gir}$ represents the random noise. 
The dependent variable $Y_{gir}$ is the amount sent by the recipient $i$ in the time interval after receiving a red packet. 
The selected time intervals are 10 minutes, 1 hour, 3 hours, 6 hours, 12 hours, and 24 hours. The main independent variable $T_{gir}$ is the amount received by user $i$ from red packet $r$. $\beta$ is the estimand of interest that specifies the linear relationship between $T_{gir}$ and $Y_{gir}$ and measures the degree of gift contagion. 
$A_{r}$, $N_{r}$,  and $O_{ir}$ refer to the total amount of the red packet $r$, its number of recipients, and user $i$'s order of receiving time, respectively. Finally, $B_s(A_{r}, N_{r}, O_{ir})$ is a dummy variable indicating whether the value of $X_{gir}=(A_{r}, N_{r}, O_{ir})$ belongs to the $s$th stratum. The dummy variable helps control for stratum fixed effects. In total, we have 180,578 strata in our sample.
%

To address potential data interdependence, we focus on group- versus user-level interdependence, as only 3.1\% of the users in our dataset belong to more than one group, data interdependence at the group level is the primary concern. To further address user-level interdependence, our bootstrap identifies any two groups containing the same user as a ``cluster.''
We use the Poisson bootstrap \citep{efron1992bootstrap} at the ``cluster'' level for 1,000 replicates to estimate the robust standard errors or 95\% confidence intervals. 

\begin{figure}
\centering
\linespread{1.25}\selectfont
\includegraphics[width=0.8\linewidth]{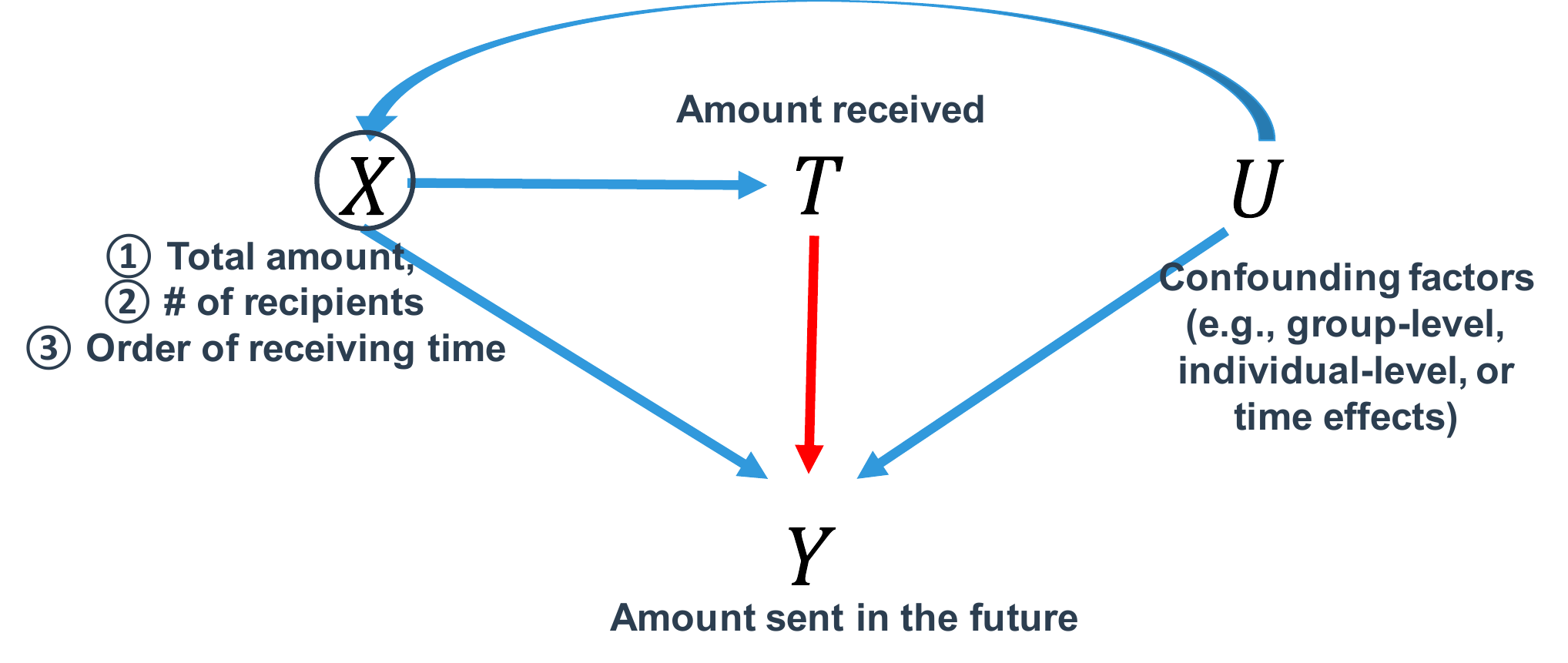}
\caption{\textcolor{black}{A directed acyclic graph illustrating the causal relationship}}
\label{fig:causal_graph}
\end{figure}

To depict the causal relationship examined by our empirical strategy, we use Pearl's directed acyclic graphs (DAGs) to visualize the causal relationship in our empirical strategy \citep{pearl2009causality}. As shown in Figure~\ref{fig:causal_graph}, controlling for $X$ blocks all of the ``backdoor'' paths from $T$ to $Y$, which satisfies the backdoor criterion and allows us to identify the causal impact of $T$ on $Y$. This process provides greater confidence that confounding factors ($U$), such as temporal clustering and homophily, would not bias our estimation. 

We also separate gift contagion into the extensive margin (the increase in the probability of sending red packets) and the intensive margin (the increase in the amount sent conditional on sending)~\citep{hossain2014crowding,liu2014crowdsourcing,bott2020you,cao2020gift}.
For the extensive margin, we replace the outcome variable in Eq.~(\ref{eq:all}) with a dummy variable $\mathbbm{1}[Y_{gir}>0]$. For the intensive margin, we apply Eq.~(\ref{eq:all}) on observations where $Y_{gir}>0$. In later analyses, we report the estimations of $\beta$ in the respective regression as extensive and intensive margins.

This empirical strategy has two advantages in identifying a causal relationship. First, it enables us to fully control for the stratum fixed effect, without requiring a specific functional form for the impact of $X$. 
For example, a linear specification, i.e., adding $A_{r}, N_{r}$, and $O_{ir}$ directly into the regression,  would lead to an overestimated treatment effect (\textit{Appendix~D.2}). 
Second, we realize that if most strata have few observations, we may fail to measure such within-stratum effects. 
Fortunately, our sample size is sufficiently large that we have a sufficient number of observations within each stratum. 
Note that the average number of observations in a stratum is 8.37. Compared to  \cite{kizilcec2018social}, which relies on the birthday discontinuity to perform a quasi-experiment, we leverage the gift-amount randomization algorithm embedded in the feature  to identify the social contagion of gift giving.


\section{Hypotheses} 
\label{sec:hyp}

To motivate the hypotheses, we construct a simple model by following~\cite{charness2002understanding}  (see \textit{Appendix~E}). 
In our model, 
the utility function of a user depends on her own payoff and that of others in the same group, with parameters specifying their respective weight.
We theoretically show that a user's sending amount increases with the amount that she receives from the preceding red packet. This leads to our first hypothesis. 

\begin{hyp}[Gift contagion]
The larger the amount a recipient obtains, the larger the amount the recipient will send to the group. 
\label{H1}
\end{hyp}

Next, inspired by the fact that individuals would exchange red packets more often during holiday seasons, we quantitatively show that the size of gift contagion is stronger during festival periods. This leads to the following hypothesis:
%

\begin{hyp}[Festival effect]
Gift contagion is stronger during festival periods than during other time periods.\label{h:festival}
\end{hyp}

Similar to the festival effect, we also predict that the strength of gift contagion would be different among different groups. In particular, we expect that the effect would be stronger in groups of relatives. This leads to the following hypothesis:


\begin{hyp}[Group type effect]
Gift contagion is stronger in groups of relatives than it is in other groups.\label{h:grouptype}
\end{hyp}

Recall that the user interface highlights the person who is the luckiest draw recipient, and this information is observed by all group members. 
We thus posit that gift contagion is stronger for the luckiest draw recipients. 
There are two possible reasons for our conjecture. The first reason is because of the \textit{amount effect}: luckiest draw recipients receive larger amounts, and thus, they may send a larger amount to others, as we explained in Hypothesis 1. 
The second reason is
that the salience of the luckiest draw recipient information might motivate users to send red packets (referred to as the \textit{luckiest draw effect}). 
In our model shown in \textit{Appendix~E}, we analytically show that such luckiest draw effect would amplify the size of gift contagion. Hence, we have the following hypothesis:

%



\begin{hyp}[Luckiest draw effect]
\textit{Gift contagion is stronger for luckiest draw recipients than it is for others.}
\label{h:luck}
\end{hyp}

Finally, we are interested in the moderating effect of social network characteristics. 
Again, we quantitatively show that a member who has more friends or who is less clustered in their group would send more, conditional on the same amount received. Therefore, we have the following hypothesis:

\begin{hyp}[Individual network position on gift contagion]{
\begin{enumerate}[label={(\alph*)}]
\item[]
\item \textit{Gift contagion is stronger for individuals who are less clustered in the group.}
\item \textit{Gift contagion is stronger for individuals who have more friends (higher degree) in the group.}
\end{enumerate}
\label{h:network}}
\end{hyp}

\section{Results}\label{sec:result}
\subsection{Gift contagion in online groups}





\begin{figure}[h!]
    \centering
\linespread{1.25}\selectfont
    \includegraphics[width=0.4\linewidth]{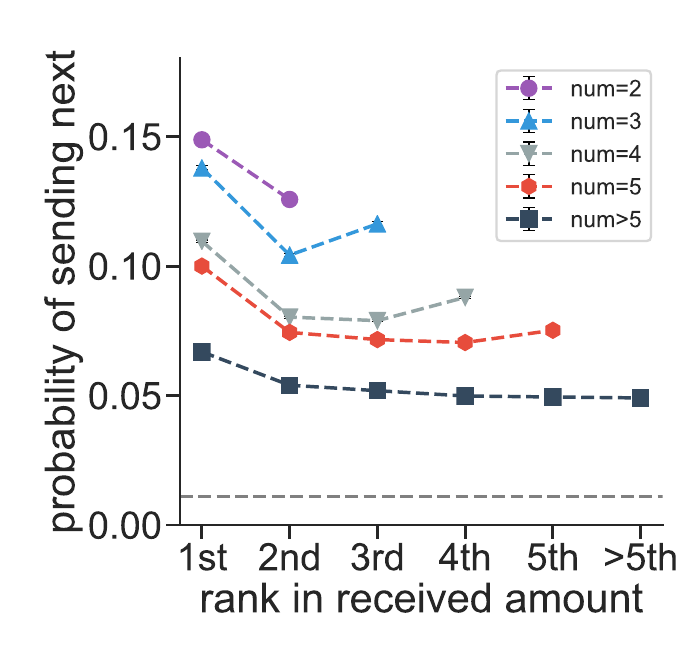}
    \caption{\raggedright \linespread{1.2} \selectfont The recipients' probability of sending the first subsequent red packet. ``Num'' is the number of recipients of a given red packet. The $x$-axis indicates the rank of received amounts among recipients. For example, ``1st'' refers to the user who receives the largest amount, i.e., the luckiest draw recipient. ``$>$5th'' is the average probability among recipients whose rank is below the 5th position. The dashed gray line represents the average probability that a non-recipient sends the first subsequent red packet. The error bars, i.e., the 95\% CIs,  are much smaller than the markers, and become invisible. }
    \label{fig:causal}
\end{figure}

We first apply a simplistic, non-parametric approach to shed light on the causal effect of the amount received by a user within a group on the probability of that user sending the first subsequent red packet. We depict the probability of sending the first subsequent red packet for the recipients of a given red packet in Figure~\ref{fig:causal}.
From this figure, we see a decreasing trend related to the rank of received amounts: those who receive the largest amount have the highest probability of sending the first subsequent red packet. Moreover, the largest difference lies between those who receive the largest amount and those who receive the second largest amount, while the differences between other recipients are much smaller.

\begin{figure}[h!]
\centering
\includegraphics[width=0.4\linewidth]{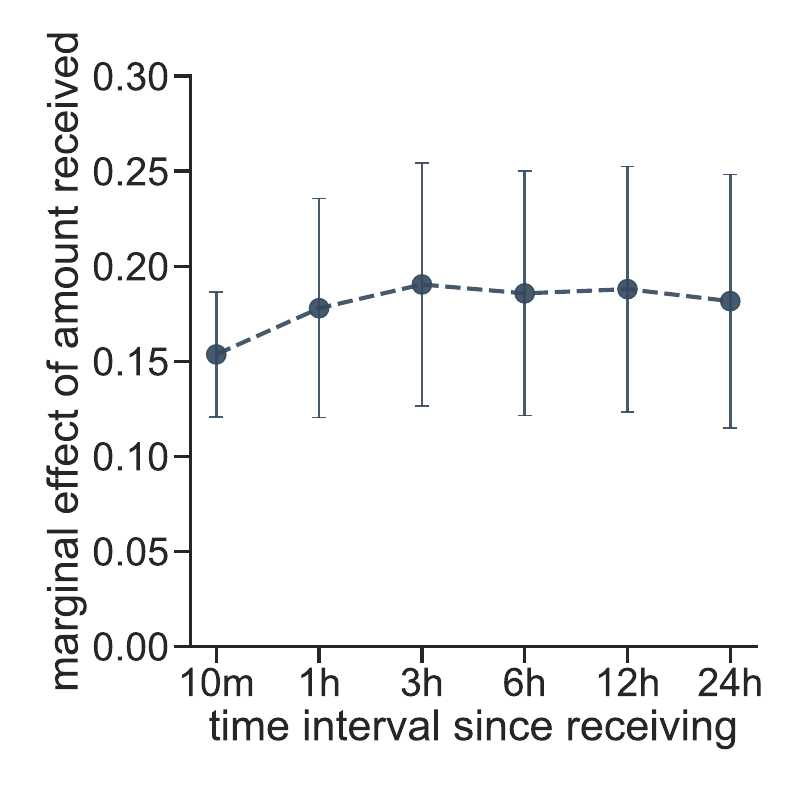}
\caption{\raggedright \linespread{1.2} \selectfont
The marginal effects of the amount received on the amount sent within the corresponding timeframes. 
Error bars are the 95\% CIs.
}
    \label{fig:summary}
\end{figure}

Next, we apply the empirical strategy described in Section~\ref{sec:strategy} to quantify the impact of the amount received on the subsequent amount sent, namely, we estimate $\beta$ in Eq.~(\ref{eq:all}). 
Figure~\ref{fig:summary} presents the marginal effects for different timeframes.
From the figure, we see an increase in the effect size as the timeframe widens, with the effect stabilizing after the first three hours. In later analyses, we focus on regression results for 10 minutes and 24 hours, respectively. 

\begin{table}[h]
\caption{Regression analyses for gift contagion}
\scriptsize
\centering
\label{tab:margin}
\begin{threeparttable}
\linespread{1.25}\selectfont
\begin{tabular}{p{3cm}cccccc}
\hline \hline
& \multicolumn{2}{c}{Overall} &
\multicolumn{2}{c}{Extensive} & \multicolumn{2}{c}{Intensive} \\
                                    & 10 min & 24 h  & 10 min & 24 h  & 10 min & 24 h   \\
& (1) & (2) & (3) & (4) & (5) & (6) 
\\  \hline 
Amount received & 0.1554*** & 0.1850*** & 0.0031*** & 0.0032*** & 0.0048 & -0.2210*\\
 & (0.0176) & (0.0359) & (0.0001) & (0.0001) & (0.0725) & (0.1374) \\

Stratum fixed effect & Y          & Y         & Y          & Y      & Y          & Y          \\ 

No. of observations                                          & 7,266,446    & 7,266,446 & 7,266,446    & 7,266,446 &    1,060,746
  & 1,370,741
    \\
Adjusted $R^2$ &  0.0394 & 0.0396 & 0.0211 & 0.0233 & 0.1517 & 0.1096 \\
\hline \hline
\end{tabular}
\begin{tablenotes}[flushleft]
\scriptsize \item \mytablenote
\end{tablenotes}
\end{threeparttable}
\end{table}

\begin{res}[Gift contagion]
\textit{The larger the amount a recipient obtains, the larger the amount she sends to the group.}
\label{res:gift}
\end{res}

\support{
\textit{As shown in Columns (1) and (2) in Table~\ref{tab:margin}, the regression coefficients for the amount received are positive and significant at the 1\% level (10 minute: 0.1554, $p<0.01$; 24-hour: 0.1850, $p<0.01$). 
}
}

By Result \ref{res:gift}, we reject the null hypothesis in favor of Hypothesis~\ref{H1}.
Since the gifting behavior is quantitatively measurable in our dataset, we are able to decompose the overall effect to the extensive and intensive margins.
This decomposition has received little examination in the previous gift contagion literature. For example, in their study of gift contagion on Facebook, \cite{kizilcec2018social} reports only the overall effect of gift contagion.
The extensive margin reflects whether the  amount received increases the recipient's likelihood of sending a red packet, while the intensive margin indicates, conditional on sending a red packet, whether the amount received affects the amount sent. 
As shown in Columns (3) and (4), receiving one more CNY increases the recipient's probability of sending a red packet by 0.31\% in 10 minutes ($p<0.01$) and 0.32\% in 24 hours ($p<0.01$). By contrast, the intensive margin is not significant within 10 minutes ($p>0.1$), and even becomes negative within 24 hours ($p=0.055$). Therefore, we conclude that the primary driver of our observed overall effect of receiving a red packet on subsequent behavior is that users are more likely to send packets versus more likely to send a greater amount.




In addition, we test for generalized reciprocity \citep{yamagishi1993generalized,nowak2007upstream}, i.e., whether receiving gifts in one group triggers the recipient to send a gift within another group.
Again, we apply the estimation strategy in Eq.~(\ref{eq:all}) to estimate the effect, but with the dependent variable being the average amount that the user sends to other groups she belongs to in our sample. Since we sample our data at the group level, our test of generalized reciprocity is restricted to those who belong to multiple sampled groups, which yields 18,910 (3.1\%) users in our sample.
Altogether, although the sign for the estimated coefficient is positive, it is not significant (see Table~A.7 in \textit{Appendix}). This null result may be due to two factors. First, although the number of users is not small, a lack of within-stratum variation may underpower our analysis. Indeed, among 12,671 strata, 7,956  contain  only one observation. Second, since users may belong to additional groups that are not in our sample, the lack of all sending and receiving history of a user leads to an underestimation of our effect.

\subsection{Heterogeneous effect of gift contagion}

The fine-grained information in our large dataset provides opportunities to examine how the effect size of gift contagion varies in multiple dimensions, which further deepens our understanding of gift contagion. For example, \citet{kizilcec2018social} only examine Facebook birthday gifts, while our sample includes the sending of red packets for different purposes, such as celebrating the Lunar New Year and job promotion.
We first examine how the strength of gift contagion differs between festivals and non-festival seasons by running regressions  separately, and report the main results below.\footnote{We consider all important dates that people celebrate in China including the Lunar New Year and other festivals.}

\begin{res}[Festival effect]
\textit{The effect of gift contagion is stronger during festival than non-festival periods, and the difference is significant in the first 10 minutes.
}
\label{res:festival}
\end{res}


\support{
As shown in Columns (1) and (2) of Table~\ref{tab:festival}, the size of the overall effect is larger during festival than non-festival seasons (10 minutes: 0.1936 versus 0.1199, $p=0.042$; 24 hours: 0.2460 versus 0.1239, $p=0.101$). 
}

By Result~\ref{res:festival}, we reject the null hypothesis in favor of Hypothesis~\ref{h:festival}. We also examine the extensive and intensive margins and present the results in Columns (3)-(6), Table~\ref{tab:festival}. 
From these results, we see that 
the extensive margin is significant for both festival and non-festival seasons, although the differences are not significant at the 5\% level (10 minutes: $p>0.1$;  24 hours: $p=0.098$). Additionally, the intensive margin for festival season is larger than non-festival season (10 minutes: $p>0.1$; 24 hours: $p=0.074$).


\begin{table}[]
\caption{Regression analyses for gift contagion: festival versus non-festival seasons}
\scriptsize
\centering
\label{tab:festival}\makebox[\linewidth]{
\begin{threeparttable}
\linespread{1.25}\selectfont
\begin{tabular}{p{4.cm}cccccccccccc}
    
\hline\hline
& \multicolumn{2}{c}{Overall} &
\multicolumn{2}{c}{Extensive} & 
\multicolumn{2}{c}{Intensive} \\
& 10 min & 24 h  & 10 min & 24 h  & 10 min & 24 h   \\  
& (1) & (2)  & (3) & (4)  & (5) &  (6)   \\ 
\hline 
& \multicolumn{6}{c}{Festival} \\
Amount received & 
0.1936*** & 0.2460*** & 0.0030*** & 0.0031*** & 0.0803 & -0.0205\\
 & (0.0261) & (0.0531) & (0.0001) & (0.0001) & (0.0831) & (0.1487) \\
Stratum fixed effect & Y & Y & Y & Y & Y & Y \\
No. of observations & 2,297,290 & 2,297,290 & 2,297,290 & 2,297,290 & 399,763 & 545,953   \\
    
Adjusted $R^2$ &  0.0458 & 0.0493 & 0.0172 & 0.0222 & 0.1626 & 0.1260 \\ \hline
& \multicolumn{6}{c}{Non-festival} \\
Amount received & 0.1199*** & 0.1239*** & 0.0032*** & 0.0034*** & -0.1571 & -0.6609** \\
 & (0.0251) & (0.0520) & (0.0001) & (0.0001) & (0.1514) & (0.3262) \\
Stratum fixed effect & Y & Y & Y & Y & Y & Y \\
No. of observations & 4,969,156 & 4,969,156 & 4,969,156 & 4,969,156 & 660,983 & 824,788  \\  
Adjusted $R^2$ & 0.0342 & 0.0318 & 0.0196 & 0.0208 & 0.1861 & 0.1485 \\ \hline
\hline
\end{tabular}

\begin{tablenotes}[flushleft]
\small
\scriptsize \item \mytablenote
\end{tablenotes}
\end{threeparttable}}
\end{table}

Second, we examine whether the gift contagion effect varies across different types of groups. We identify three group types by inferring a group's composition from group names. 
\begin{CJK*}{UTF8}{gbsn}
(1) \textit{Relative groups}: groups with names containing 家(family).
(2) \textit{Classmate groups}: groups with names containing [班(class), 小学/中学/初中/高中(elementary/secondary/low secondary/high secondary school,
respectively), 大学(college/university), 校(school), 届/级(grade)].
(3) \textit{Coworker groups}: groups with names containing 公司(company), 集团(corporate group), 工作(work), and 有限(limited liability).
\end{CJK*} 
Table~\ref{tab:grouphetero} reports the regression results for group type analysis. 
%

\begin{table}[]
\caption{Regression analyses for gift contagion by group types}
\scriptsize
\centering
\label{tab:grouphetero}\makebox[\linewidth]{
\begin{threeparttable}
\linespread{1.25}\selectfont
\begin{tabular}{p{4.cm}cccccc}
\hline\hline
& \multicolumn{2}{c}{Overall} &
\multicolumn{2}{c}{Extensive} & 
\multicolumn{2}{c}{Intensive} 
\\
& 10 min & 24 h  & 10 min & 24 h  & 10 min & 24 h\\  
& (1) & (2)  & (3) & (4)  & (5) &  (6)   \\  \hline 

& \multicolumn{6}{c}{Relatives} \\
Amount received & 0.1484*** & 0.1948*** & 0.0031*** & 0.0031*** & 0.0792 & 0.0305\\
 & (0.0227) & (0.0443) & (0.0002) & (0.0002) & (0.0816) & (0.1298) \\
Stratum fixed effect  & Y  & Y  &  Y & Y  & Y  & Y  \\

No. of observations &  2,200,404 & 2,200,404 & 2,200,404 & 2,200,404 & 366,553 & 472,239 \\
Adjusted $R^2$ &   0.0636 & 0.0552 & 0.0139 & 0.0169 & 0.2169 & 0.1405 \\ \hline
& \multicolumn{6}{c}{Classmates}  \\
Amount received & -0.0253 & -0.0913 & 0.0068*** & 0.0069*** & -1.1597** & -1.2252**\\
 & (0.0625) & (0.1245) & (0.0009) & (0.0009) & (0.4202) & (0.5316) \\
Stratum fixed effect  & Y  & Y  &  Y  &  Y  & Y  & Y   \\ 
No. of observations & 408,397 & 408,397 & 408,397 & 408,397 & 47,242 & 62,616
    \\
Adjusted $R^2$ &   0.0982 & 0.1206 & 0.0082 & 0.0148 & 0.2631 & 0.1956 \\ \hline 
& \multicolumn{6}{c}{Coworkers} \\
Amount received & 0.1267** & 0.0627 & 0.0032*** & 0.0031*** & -0.1841 & -0.6356\\
 & (0.0643) & (0.1062) & (0.0005) & (0.0006) & (0.3836) & (0.5495) \\
Stratum fixed effect  & Y  & Y  &  Y & Y  & Y  & Y \\ 

No. of observations            & 143,297 & 143,297 & 143,297 & 143,297 & 17,974 & 23,156                              
    \\
Adjusted $R^2$ &  0.1633 & 0.1723 & -0.0067 & 0.0041 & 0.3694 & 0.3236   \\

\hline \hline
\end{tabular}
\begin{tablenotes}[flushleft]
\item \scriptsize \mytablenote
\end{tablenotes}
\end{threeparttable}}
\end{table}

%

\begin{res}[Group type effect]
The overall effect of gift contagion is  stronger in relative groups than in classmate groups. 
\label{res:grouptype}
\end{res}

\support{
The effect size is significantly greater in relative groups than in classmate groups (10 minutes: $0.1484$ versus $-0.0253$, $p=0.009$; 24 hours: $0.1948$ versus $-0.0913$, $p=0.031$). 
}

By Result~\ref{res:grouptype}, we reject the null hypothesis in favor of Hypothesis~\ref{h:grouptype}. In addition, as indicated by Table~\ref{res:grouptype}, we find that the overall effect is significant for relative groups, while the effects are not significant for the other two groups.\footnote{The overall effect is marginally significant for coworker groups only within 10 minutes ($p<0.1$).} Moreover, we find that the effect on extensive margins is positive and significant across all group types, but not on intensive margins.\footnote{Surprisingly, the effect on intensive margin is even negative and significant for classmates groups.} This suggests that the effect of gift contagion is primarily driven by promoting new participants to join the red packet chain.
These results are consistent with the prior literature, which shows that the overall effect on workers' productivity is primarily driven by the extensive margin rather than the intensive margin~\citep{hossain2014crowding,cao2020gift}.
Additionally, we compare the effect size between group types and find that the extensive margin is significantly higher in classmate groups than in relative groups (10 minutes: 0.0068 versus 0.0031, $p<0.01$; 24 hours: 0.0069 versus 0.0031, $p<0.01$) or coworker groups (10 minutes: 0.0068 versus 0.0032, $p<0.01$; 24 hours: 0.0069 versus 0.0031, $p<0.01$). By contrast, the intensive margin for classmate groups is significantly smaller than that for relative groups (10 minutes: $-1.1597$ versus $0.0792$, $p<0.01$; 24 hours: $-1.2252$ versus 0.0305, $p=0.022$). 

%

In addition, we investigate how other demographic characteristics, such as gender and age, affect the degree of gift contagion. We find that the effect of gift contagion is stronger for older recipients, and that red packets sent by younger users are more socially contagious. Moreover, although there is no significant gender difference in the overall effect, red packets sent by female users tend to exhibit a higher extensive margin. \textit{Appendix~D.2} includes detailed analyses.

\subsection{``Luckiest draw'' effect }
As discussed in Section~\ref{sec:hyp}, we posit that luckiest draw recipients may exhibit stronger gift contagion. 
To examine the behavioral difference between luckiest and non-luckiest draw recipients, we run the regressions in Eq.~(\ref{eq:all}) for these two subgroups separately and report the results in Table~\ref{tab:luckiest}.

\begin{res}[Luckiest draw effect]
\textit{Gift contagion is stronger for luckiest draw recipients than non-luckiest draw recipients, and the difference is significant in the 10-minute timeframe.  
}
\label{res:luck}
\end{res}

\support{
In Columns (1) and (2) of Table~\ref{tab:luckiest}, the marginal effects for luckiest draw recipients are larger than those for non-luckiest draw recipients (10 minutes: 0.3271 versus 0.0981, $p=0.011$; 24 hours: 0.3985 versus 0.1616, $p>0.1$). 
}

By Result~\ref{res:luck}, we reject the null hypothesis in favor of Hypothesis~\ref{h:luck}. 
Moreover, we find that the extensive margin for luckiest draw recipients is almost ten times of that for non-luckiest draw recipients (10 minutes: $0.0074$ versus $0.0007$, $p<0.01$; 24 hours: $0.0078$ versus $0.0008$, $p<0.01$).  Conditional on sending red packets, the marginal effect on the amount that a user sends is smaller for luckiest than non-luckiest draw recipients, although the difference is only marginally significant (Columns (5) and (6), 10 minutes: $-0.3767$ versus $0.2246$, $p=0.064$; 24 hours: $-0.8846$ versus $0.2030$, $p=0.067$). 

\begin{table}[h]
\caption{Regression analyses for gift contagion: luckiest versus non-luckiest draw recipients }
\scriptsize
\centering
\label{tab:luckiest}\makebox[\linewidth]{
\begin{threeparttable}
\linespread{1.25}\selectfont
\begin{tabular}{p{4.cm}cccccc}
\hline\hline
& \multicolumn{2}{c}{Overall} &
\multicolumn{2}{c}{Extensive} & 
\multicolumn{2}{c}{Intensive} 
\\ & 10 min & 24 h  & 10 min & 24 h  & 10 min & 24 h \\  
& (1) & (2) & (3) & (4) & (5) & (6) \\
\hline 
& \multicolumn{6}{c}{Luckiest} \\
Amount received & 0.3271*** & 0.3985*** & 0.0074*** & 0.0078*** & -0.3767 & -0.8846*\\
 & (0.0836) & (0.1635) & (0.0004) & (0.0004) & (0.2743) & (0.4862) \\
Stratum fixed effect & Y & Y & Y & Y & Y & Y    \\ 
No. of observations & 1,923,297 & 1,923,297 & 1,923,297 & 1,923,297 & 296,799 & 371,698  \\
Adjusted $R^2$ &0.0640 & 0.0503 & 0.0348 & 0.0373 & 0.1844 & 0.1222 \\
\hline 
& \multicolumn{6}{c}{Non-luckiest} \\
Amount received & 0.0981*** & 0.1616** & 0.0007*** & 0.0008*** & 0.2246 & 0.2030\\
 & (0.0331) & (0.0807) & (0.0001) & (0.0001) & (0.1720) & (0.3393) \\
Stratum fixed effect & Y & Y & Y & Y & Y & Y  \\ 
No. of observations & 5,343,149 & 5,343,149 & 5,343,149 & 5,343,149 & 763,947 & 999,043  \\
Adjusted $R^2$ & 0.0373 & 0.0413 & 0.0159 & 0.0184 & 0.1543 & 0.1181  \\\hline  \hline
\end{tabular}
\begin{tablenotes}[flushleft]
\item \scriptsize \mytablenote
\end{tablenotes}
\end{threeparttable}}
\end{table}


It is possible that luckiest draw recipients send more simply because they receive more. To control for this   ``amount'' effect, we implement the following matching procedure. 
Specifically, we match each luckiest draw recipient with non-luckiest draw recipients by holding the following variables constant: the total amount of the red packet, the number of recipients of that red packet, the order of receiving time, and the amount received by the corresponding recipient. Matching on the first three variables allows us to control for the effect of unobserved variables, as the backdoor criterion is satisfied \citep{pearl2009causality}. Moreover, matching on the received amount allows us to further control for the difference in the amount received.\footnote{We conduct  one-to-many matching \citep{stuart2010matching}.} 
Our matching procedure yields 668,936 luckiest draw recipients and 1,658,283 non-luckiest draw recipients, representing successful matching of 33.7\% of our luckiest draw recipients. Additionally, we bootstrap for 1,000 replicates to construct the confidence intervals.


\begin{figure}[h]
\includegraphics[width=0.95\linewidth]{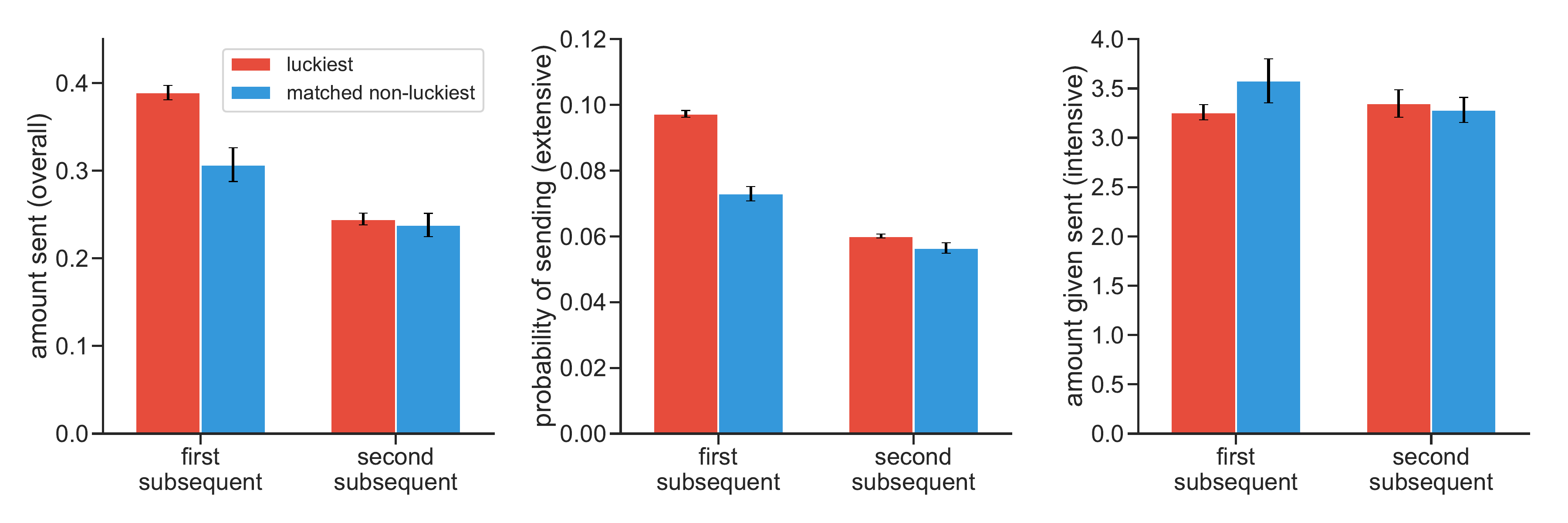}
\caption{\raggedright \linespread{1.2} \selectfont Comparisons between luckiest and non-luckiest draw recipients for the unconditional amount that a user sends ((left), the probability of sending (middle), and the conditional amount that a user sends (right). 
{Error bars are the 95\% CIs.}
}
\label{fig:luck_match}
\end{figure}


As shown in the left panel of Figure~\ref{fig:luck_match}, the cash amount sent in the first subsequent red packet increases by 0.080 CNY from non-luckiest to luckiest draw recipients, an effect that is significant at the 1\% level. This result suggests that
being the luckiest draw recipient alone promotes the gift contagion. By contrast, for the \textit{second} subsequent red packet,  we find a much smaller increase and the effect is no longer significant  ($p>0.1$).\footnote{There is no significant difference for the third and subsequent red packets.} 
We also decompose the overall effect into the extensive and intensive margins.\footnote{Note that the definitions here are slightly different: (1) overall: the amount sent in the $k$th subsequent red packet; (2) extensive margin: whether the recipient sent the  $k$th subsequent red packet; and (3) intensive margin: the amount sent conditional on being the user who sends the $k$th subsequent red packet.} For both the first and second subsequent red packets, we find significant differences for the extensive margins ($p<0.01$), although the difference for the first subsequent red packets is much larger. For the intensive margin, the effect of luckiest draw recipients is smaller than non-luckiest draw recipients for the first subsequent ($p=0.013$) but not for the second subsequent red packet ($p>0.1$). 




Our results suggest the existence of a group norm whereby the luckiest draw recipients should send the first subsequent red packet. This norm can facilitate coordination among group members to maintain a chain of red packets \citep{feldman1984development,seinen2006social,gachter2013peer}.  
Moreover, we find that the strength of such a group norm is contingent on the discrepancy between the amounts received by the luckiest draw recipient and the amounts received by other recipients  (\textit{Appendix~D.2}), suggesting that the fairness concern plays a role in influencing the strength of the luckiest draw effect \citep{bolton2000erc,bolton2006inequality}. 
In addition to the group norm, gifts may pressure recipients into signaling their own virtue, especially for luckiest-draw recipients. However, since users endogenously decide whether they want to receive a red packet, they can avoid social pressure by not clicking on an offered packet. In addition, the upper limit of a red packet’s cash amount is not very large -- 200 CNY (roughly 30 USD) and the average cash amount for luckiest-draw recipients in our setting is 1.16 CNY. Therefore, we suspect that social pressure or reputational concerns do not play an important role in our setting. 

\subsection{Social contagion and social network}

In this section, we apply social network analysis to understand how the group network structure affects the strength of our observed gift contagion. 
Since group members may or may not be private contacts (``friends''), we construct a relationship network among group members, with each edge indicating that two group members are contacts. 


First, we examine how individual network positions affect gift contagion. Specifically,we focus on the clustering coefficient and degree. 
The clustering coefficient of user $i$ in group $g$ \citep{holland1971transitivity,watts1998collective}, or the extent to which a user's friends are connected, is defined below:

\begin{equation}
\text{clustering coefficient}(i,g) = 
\frac{\sum_{j \in \mathcal{N}_i^g} \sum_{k \in \mathcal{N}_i^g, k \neq j} \mathbbm{1} [k \in \mathcal{N}_j^g]}{|\mathcal{N}_i^g|(|\mathcal{N}_i^g| - 1)},
\end{equation}

\noindent where $\mathcal{N}_i^g$ is the set of a user $i$'s in-group friends in group $g$.  
 The value of the clustering coefficient ranges from $[0, 1]$; 0 indicates that none of $i$'s friends are connected and 1 indicates that all of $i$'s friends are connected in a group. 
 Moreover, we use the normalized degree in our analysis: $\frac{\text{degree}(i,g)}{\text{No. of group members}}$, with a range of $[0, 1]$.\footnote{Compared to the (unnormalized) degree, normalized degree considers the effect of group size.}
 
Table~\ref{tab:degreeclustering} reports the regression results adding the clustering coefficient, the normalized degree, and their interaction terms with the amount received as independent variables. We summarize the results below:

\begin{res}[Individual network position on gift contagion]
\begin{enumerate}[label={(\alph*)}]
\item []
\item The overall effect of gift contagion is smaller for group members with a higher clustering coefficient. 
\item The normalized degree does not significantly impact the overall effect of gift contagion.
\end{enumerate}
\label{res:central}
\end{res}

\support{
As shown in Columns (1) and (2) of Table~\ref{tab:degreeclustering}, the interaction term between ``Amount received'' and ``Clustering coefficient'' is negative and significant at the 1\% level (10 minutes: $-0.2918$, $p<0.01$; 24 hours: $-0.7596$, $p<0.01$), and the clustering coefficient itself is also negative and significant (10 minutes: $-0.1237$, $p<0.05$; 24 hours: $-0.4350$, $p<0.01$).  
The coefficient for normalized degree is positive and significant at the 1\% level (10 minutes: $1.0236$, $p<0.01$; 
24 hours: $2.3044$, $p<0.01$), although its interaction term with ``Amount received'' is not significant.
}

By Result~\ref{res:central}(a), we reject the null hypothesis in favor of Hypothesis~\ref{h:network}(a). This finding is consistent with prior studies  \citep{aral2012identifying,ugander2012structural}.  Moreover, as shown in Columns (3)-(6), the interaction terms for the extensive and intensive margins are also  negative and significant.\footnote{The only exception is the intensive margin result for 10 minutes.} 
For the normalized degree, we do not find a salient interaction effect, and thus we fail to reject the null hypothesis in favor of Hypothesis~\ref{h:network}(b).\footnote{{We also examine the impact of centrality, in particular, the eigenvector centrality, which is widely used in the literature of networks \citep{marsden2002egocentric,jackson2010social}. However, no significant overall effect for the interaction term   is found  (Table~A.13).} }

\begin{table}[h]
    \centering
    \caption{Effect of individual in-group degree and clustering coefficient on gift contagion }
        \label{tab:degreeclustering}
    \linespread{1.25}\selectfont
    \scriptsize\makebox[\linewidth]{
    \begin{threeparttable}
    \begin{tabular}{lcccccc}
    \hline \hline
    	& \multicolumn{2}{c}{Overall} 	&  	\multicolumn{2}{c}{Extensive}  &  \multicolumn{2}{c}{Intensive}	\\
    	& 10 min	& 24 h	& 10 min   &	24 h &   10 min &   24 h \\
    	& (1) & (2) & (3) & (4) & (5) & (6) \\  \hline
    Amount received &	0.3456*** &	0.6748*** &	0.0084*** &	0.0088*** &	0.1261 &	0.4101 \\
    	 & (0.0858)  &	(0.2084)  &	(0.0004)  &	(0.0005)  &	(0.3030)  &	(0.5835) \\
    Amount received $\times$ normalized degree  &	0.0674	 & 0.1839  &	-0.0051***  &	-0.0053***  &	0.3049  &	0.2794 \\
    	& (0.0910)  &	(0.2058)  &	(0.0004)  &	(0.0004)  &	(0.3515) &	(0.6398) \\
    Amount received $\times$ clustering coefficient & -0.2918*** &	-0.7596*** &	-0.0028*** &	-0.0029*** &	-0.3899 &	-1.0140** \\
    	& (0.0742) &	(0.1594) &	(0.0004)  &	(0.0004)  &	(0.2492)	 & (0.4404) \\
    Normalized degree &	1.0236*** &	2.3044*** &	0.0656*** &	0.0905*** &	3.8504*** &	6.7114*** \\
     &	(0.0710) &	(0.1586) &	(0.0012) &	(0.0014) &	(0.3725)	& (0.6468) \\
    Clustering coefficient&	-0.1237** &	-0.4350*** &	-0.0438*** &	-0.0636*** &	1.6890*** &	2.1104*** \\
    &	(0.0519) &	(0.1101) &	(0.0010) &	(0.0012) &	(0.2706) &	(0.4650) \\
    Group size & Y & Y & Y & Y & Y & Y \\
    Stratum fixed effect & Y & Y & Y & Y & Y & Y \\ 
    No. of observations	& 7,266,446 &	7,266,446 &	7,266,446	& 7,266,446	 & 1,060,746 &	1,370,741\\
    Adjusted $R^2$ &	0.0400 &	0.0403	&  0.0260	& 0.0308	& 0.1524	& 0.1102  \\
     \hline  \hline
    \end{tabular}
    \begin{tablenotes}[flushleft]
     \scriptsize \item \mytablenote
    \end{tablenotes}
    \end{threeparttable}}
\end{table}

Next, we examine the effect of the group-level network structure on our observed gift contagion. 
We use \textit{the average normalized degree}, or \textit{network density} to measure the degree to which a network is tightly connected \citep{newman2006structure}:

\begin{equation}
\text{average normalized degree}(g) = 
\frac{ \sum_{i \in \mathcal{G}} |\mathcal{N}_i^g|}{ |\mathcal{G}| \times (|\mathcal{G}|-1) }.
\end{equation}

\noindent $\mathcal{G}$ denotes the set of group $g$'s members. The average normalized degree ranges from [0, 1]. We present results of the regression with average normalized degree and the interaction term in Table~\ref{tab:density}.

\begin{table}[h]
    \centering
    \caption{Effect of average normalized degree in groups }
        \label{tab:density}
    \linespread{1.25}\selectfont
    \scriptsize\makebox[\linewidth]{
    \begin{threeparttable}
    \begin{tabular}{lcccccc}
    \hline \hline
    	& \multicolumn{2}{c}{Overall} 	&  	\multicolumn{2}{c}{Extensive}  &  \multicolumn{2}{c}{Intensive}	\\
    	& 10 min	& 24 h	& 10 min   &	24 h &   10 min &   24 h \\
    	& (1) & (2) & (3) & (4) & (5) & (6) \\  \hline
        Amount received & 0.2165*** & 0.4092*** & 0.0062*** & 0.0062*** & 0.0144 & 0.1839 \\
        & (0.0578) & (0.1305) & (0.0003) & (0.0003) & (0.2024) & (0.3737) \\
        Amount received $\times$ avg normalized degree & -0.0902 & -0.3313 & -0.0046*** & -0.0045*** & -0.0121 & -0.5867 \\
        & (0.0868) & (0.2035) & (0.0004) & (0.0004) & (0.2983) & (0.5848) \\
        Avg normalized degree & 0.8661*** & 1.7084*** & 0.0184*** & 0.0192*** & 4.7967*** & 7.1958*** \\
        & (0.0806) & (0.1819) & (0.0019) & (0.0022) & (0.4218) & (0.7421) \\
        Group size & Y & Y & Y & Y & Y & Y \\
        Stratum fixed effect & Y & Y & Y & Y & Y & Y \\ 
    No. of observations	& 7,266,446 &	7,266,446 &	7,266,446	& 7,266,446	 & 1,060,746 &	1,370,741\\
    Adjusted $R^2$ & 0.0397 & 0.0399 & 0.0239 & 0.0272 & 0.1523 & 0.1100 \\
     \hline  \hline
    \end{tabular}
    \begin{tablenotes}[flushleft]
     \scriptsize \item \mytablenote
    \end{tablenotes}
    \end{threeparttable}}
\end{table}

We find that although there is no significant overall impact of group network structure on gift contagion, the interaction effect between the amount received and the average normalized degree is negative and significant for the extensive margin. 
%
%
%
%
In Columns (3) and (4) of Table \ref{tab:density}, the interaction term for ``Amount received $\times$ avg normalized degree'' is negative and significant (10 minutes: $-0.0046$, $p<0.01$; 24 hours: $-0.0045$, $p<0.01$), although there is no significant interaction effect for overall effects or intensive margins.
We also examine the impact of overall clustering \citep{jackson2010social} and find similar results. The detailed analyses are reported in \textit{Appendix~D.2}.

\begin{figure}[h!]
    \centering
    \includegraphics[width=0.4\linewidth]{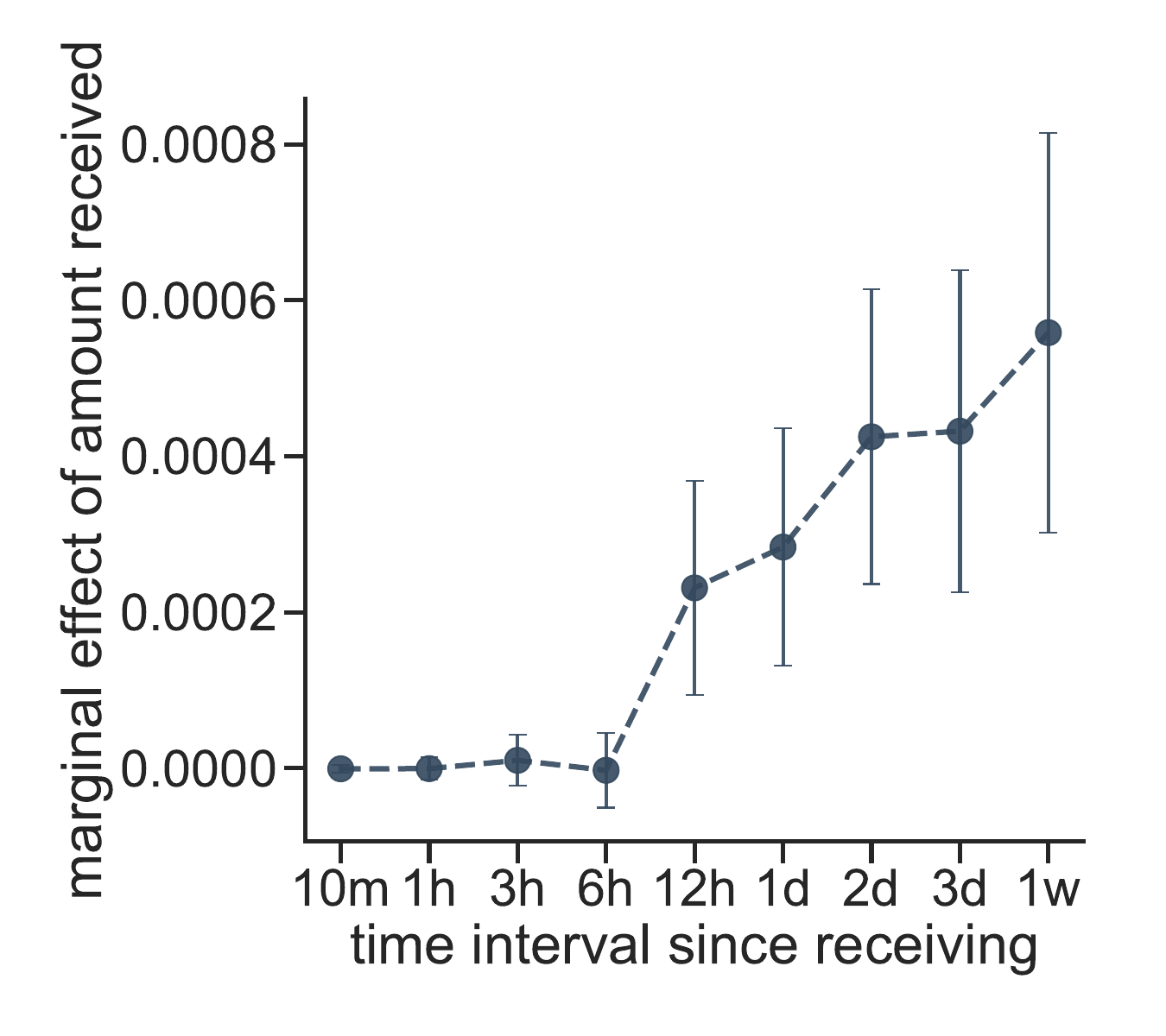}
    \caption{\raggedright \linespread{1.2} \selectfont The marginal effect on the within-group edges added by the recipient within the group. Error bars are the 95\% CIs.}
    \label{fig:dynamic}
\end{figure}

Finally, we examine the impact of receiving gifts  on network dynamics. 
We change the dependent variable in Eq.~(\ref{eq:all}) to the number of within-group edges added by a user after the user receives a red packet. 
Figure~\ref{fig:dynamic} presents the results, where the $x$-axis indicates different time intervals and the $y$-axis represents the marginal effect of the amount received (in CNY) on the number of new friends added by the recipient within the group. 
On average, receiving 100 CNY encourages the recipient to add 0.055 friends within the group in the subsequent seven days ($p<0.01$). Although this appears to be a small effect, it reflects how in-group gifts can foster in-group interactions through establishing new connections. 
In sum, our findings suggest that in-groups gifts not only promote gift contagion, but can also encourage within-group interaction and strengthen group solidarity.

\section{Discussion}\label{sec:dis}

Taking advantage of the random assignment of red packet amounts to gift recipients, we leverage a natural experiment to quantify the strength of gift contagion within online groups. We document the presence of gift contagion and further find that the overall effect is driven primarily by the extensive margin, i.e., receiving red packets encourages more users to send packets. The degree of gift contagion varies across different time periods and various groups. Moreover, we find evidence of a group norm whereby the luckiest draw recipients are expected to take the lead in sending the first subsequent red packet. Regarding the moderating effect of in-group social networks, we find that the higher a user’s clustering coefficient is, the less susceptible she is to gift contagion. Additionally, there is a significantly negative interaction effect for the extensive margin between the amount received and how tightly knit a  group network is. Overall, our results, especially the analyses for the extensive and intensive margins, deepen our understanding of the social phenomenon of gift contagion.

{Our study has important managerial implications. First,  
 online group chats facilitate communication and coordination, but the virtual format may pose challenges to establishing group solidarity. 
 Our results show that receiving larger amounts  not only promotes the recipient to send more gifts subsequently, but also encourages the recipient to add more friends within the group.
 These results offer insights into how one might utilize online gift giving to foster social bonds within both online groups and offline groups.}
Moreover, because of the popularity of online gifts and online red packets across different platforms,\footnote{Examples of this include Pinduoduo, the largest group buying platform in China; Alipay, the largest online payment platform in China; and other Asian bank apps such as DBS Bank and OCBC Bank in Singapore.} our heterogeneity analyses offer insights into how to leverage gift contagion to promote the adoption of online gifts. 
Because of the stronger gift contagion that occurs during festival periods, we advise that marketing campaigns be conducted during those periods. 
In addition, since we find that gift contagion is stronger in groups of relatives, platforms are advised to make additional efforts to improve the gift design for relatives and family members, which may further promote the adoption of online gifts through these social relationships.
Finally, the finding that the strength of gift contagion varies with social network characteristics helps us understand whom the marketing campaigns should be concentrated on gift production promotion.

There are several possible future directions based on our study. First, it would be interesting to examine how receiving red packets affects other types of user behaviors, such as group communication and liking others’ feeds. Second, due to data constraints, we are not able to disentangle which mechanism, such as reciprocity or fairness concern, is the main driver for our observed gift contagion. Therefore, carefully design experimental studies are needed for future work to investigate the primary mechanism. Finally, as we are analyzing online gift contagion in East Asian culture, it would be interesting to explore whether our results can be generalized to offline settings or other culture groups.

\ACKNOWLEDGMENT{For helpful discussions and comments, the authors thank Sinan Aral, Dean Eckles, Moshe Hoffman, and Erez Yoeli, as well as conference attendees at Advances in Field Experiments, Network Science in Economics, MIT Conference on Digital Experimentation, and International Conference on Computational Social Science. The authors also thank Shu Wang for his excellent research assistance.  Liu gratefully acknowledges financial support by NSFC (72222005) and Tsinghua University (No. 2022Z04W01032). Tang is supported by NSFC for Distinguished Young Scholar (61825602). Yuan and Pentland greatly acknowledge MIT Connection Science Consortium. 

}

\setcounter{section}{0} 
\setcounter{figure}{0} 
\setcounter{table}{0}    

\renewcommand\theHsubsection{\Alph{section}.\arabic{subsection}}    
\renewcommand\thesection{\Alph{section}}    
\renewcommand\theHsection{\Alph{section}}    
\renewcommand\theHfigure{A.\arabic{figure}}    
\renewcommand\theHtable{A.\arabic{table}}

\section{A Brief History of Red Packets}

\label{sec:history}

Red packets are typically 
sent from older relatives to children or unmarried young people. Children and unmarried young people wish their older relatives a ``Happy New Year,'' which is called \textit{bainian} (\begin{CJK*}{UTF8}{gbsn}拜年\end{CJK*}). Then, the older relatives give them cash gifts in exchange. Red packets are a traditional custom dating back to the Han dynasty (circa 50 BC - 100 AD) \citep{siu2001red}. 

Lucky money 
was once called ``\begin{CJK*}{UTF8}{gbsn}压祟钱\end{CJK*}'', the literal meaning of which is the money that drives away the demon \textit{Sui} (\begin{CJK*}{UTF8}{gbsn}祟\end{CJK*}) on the Lunar New Year’s Eve \citep{roy2005traditional}. ``\begin{CJK*}{UTF8}{gbsn}岁\end{CJK*}'' (age) has the same pronunciation as ``\begin{CJK*}{UTF8}{gbsn}祟\end{CJK*}.'' Sui was a demon who would enter houses on New Year Eve’s and deliberately terrify and harm children. Children would catch a terrible fever and even mental disorders if they became terrified. To protect the children, parents gave as an offering eight copper coins wrapped in red packets. It was believed that these eight coins would emit strong lights that would drive the demon away. These eight copper coins were considered the initial version of red packets. 

In the 1900s, when the printing technique was popularized in China, red packets have been developed into their current form. Chinese characters symbolizing good wishes are printed on red packets. 
Red packets are no longer used only to ensure the safety of children for superstitious reasons. 
At present, they  usually symbolize senders’ wishes for successful fortune, health, studies, and career paths. The wrapping of red packets typically contains characters with such meanings. 

In addition to the role of wishes for children or unmarried young people, red packets are also used as cash gifts on other occasions. Invitees to birthday parties, weddings, and funerals are expected to bring cash gifts, usually wrapped in red envelopes to the hosts. The amount of the gift represents the senders’ evaluation of the strength of social bonds and relationships between senders and recipients. Receiving cash gifts is regarded as owing ``\textit{renqing}'' (favor), and recipients are strongly expected to send back the gift cash in the future (called to \textit{return ``renqing''}) \citep{wang2008significance,bulte2018forced}. 


At present, with the proliferation of online platforms, red packets are commonly sent on these platforms. On the platform we study, red packets are used as convenient cash gifts, through either one-to-one or one-to-many channels. These online red packets are no longer only sent from older people to younger people, nor are they only used for the Lunar New Year or important events. The limit of these red packets (typically 200 CNY or approximately 30 USD) reduces the potential social pressure to   reciprocate with large-amount cash gifts. Users even use them for entertainment.
On other online platforms, red packets are a means of providing coupons to users. Incorporating good wishes for the customers, these red packets may encourage consumption and user engagement.

\section{Sample Description}

\label{sec:si:data}

Our sample includes 3,450,540 unique users in 174,131 groups. For each user, we obtain the demographic information listed below. For variables that are updated monthly, we use the information retrieved in February 2016 for our analysis. Moreover, we identify friendships, i.e., whether users are contacts, between users in our dataset. We summarize our data below:

\begin{itemize}
\item Group 
\begin{itemize}
\item Group size: the number of group members in a group.
\end{itemize}
\item Group members 
\begin{itemize}
\item Gender: self-reported by users.
\item Age: provided by the platform.
\item Number of groups that a user joins.
\item Number of private contacts (``friends'') that a user has on the platform.
\item Within-group degree: number of private contacts (or ``friends'') that a user has in one group. Note that it is possible that members of the same group may not be ``friends.''
\item Clustering coefficient: the extent to which a user's friends are connected in the group, as defined in the main text.

\end{itemize}
\item Red packet sending variables 

\begin{itemize}
\item Sending time.
\item Sender ID.
\item Total cash amount of the red packet, determined by the sender.
\item The number of recipients, determined by the sender.
\end{itemize}
\item Red packet receiving variables 
\begin{itemize}
\item Recipient ID.
\item Receiving time. The time interval between a red packet being sent and being received by the current recipient. A red packet expires 24 hours after being sent. We use the receiving time to infer the order of receiving time of a given red packet $r$.
\item The cash amount received.
\end{itemize}
\end{itemize}


We report the summary statistics for group size, the total number of red packets, and the total cash amount of red packets for each group in Table~\ref{tab:group}. 
We also present summary statistics for users' gender and age in Table~\ref{tab:user1}. Those between the ages of 20 and 30 represent a large proportion of our sample. In Table~\ref{tab:user2}, we further report information for within-group degree (how many private contacts, or ``friends,''  a user has in a group), the number of private contacts, and the number of groups that she joins on the platform.

Finally, we summarize information on red packets (Table~\ref{tab:rp}), including the cash amount, the number of recipients, the time interval between two successive red packets in a group, and the total completion time. We find that most red packets contain relatively small amounts (75\% of them do not exceed 5 CNY). In addition, the time intervals between two successive red packets are generally small, with all of the money from a given red packet often being received within minutes.

\section{Randomization Check}

\label{sec:si:random}
Conditional on the three variables that determine our stratification, we show that the received amount ($T$) is independent of the following variables:
(1) whether the user is \texttt{female}; 
(2) the user's \texttt{age}; 
(3) within-group \texttt{degree}, or the number of ``friends'';
(4) the number of friends on the platform (denoted by \texttt{fricnt}); 
(5) the number of groups that the user joins (denoted by \texttt{joincnt}); 
(6) the total amount of red packets that the user has sent in the group (denoted by \texttt{history\_sendamt}); 
(7) the total number of red packets that the user has sent in the group (denoted by \texttt{history\_sendcnt}); 
(8) the total amount of red packets that the user has received in the group (denoted by \texttt{history\_recvamt}); 
(9) the total number of red packets that the user has received in the group (denoted by \texttt{history\_recvcnt});
(10) the total amount of red packets sent in the group by all group members historically (denoted by \texttt{groupamt}); 
and (11) the total number of red packets sent in the group by all group members historically  (denoted by \texttt{groupnum}).

Specifically, we run simple OLS regressions for each stratum in which the dependent variable is one of the aforementioned variables, and the independent variable is the cash amount received by a user. We present the corresponding coefficients and the adjusted $p$-values after implementing false discovery control\footnote{We use the Benejamini-Hochberg procedure with $\alpha=0.1$ because it is more conservative and generates smaller adjusted $p$ values than methods such as the Bonferroni correction.}  for the two representative cases in Tables~\ref{tab:prandom1} and \ref{tab:prandom2}, respectively. 
In summary, no significant correlation is found. We also check other combinations of the amount sent and the number of recipients, and no significance is found. Overall, our data pass the randomization check. 

\subsection*{Calculation of the variance of the amount received}
\label{sec:variance}

Here we provide a complete calculation for the variance of the amount received by the $o$th recipient.  Although the expected amount is the same for different recipients, we show that their variance is generally different. Specifically, we show a non-decreasing trend for the variance with $o$. Let $S_o$ denote the summation of the first $o$ recipients' amounts received ($S_o = V_1 + V_2 + ... V_o$). Recall that $a$ is the total amount of the red packet,  $n$ is the number of recipients, and $V_o$ is the amount received by the $o$th recipient. 

We first consider the case in which $o<n$:
\begin{equation}
\begin{split}
\mathbbm{E}[S_{o+1}^2] & = \mathbbm{E}\big[(S_o+V_{o+1})^2\big]  =  \mathbbm{E}[S_{o}^2] + 2 \mathbbm{E} \Big[S_o \times \frac{a-S_o}{n-o}\Big] + \frac{4}{3} \mathbbm{E} \Big[ \frac{(a-S_o)^2}{(n-o)^2} \Big]  \\
& = \mathbbm{E}[S_o^2] \Big(1  - \frac{2}{n-o} + \frac{4}{3} \frac{1}{(n-o)^2}\Big) + \mathbbm{E}[S_o] \bigg(\frac{2a}{n-o} - \frac{8}{3} \frac{a}{(n-o)^2} \bigg) + \frac{4}{3} \frac{a^2}{(n-o)^2}  \\ 
& = \mathbbm{E}[S_o^2] \Big(1  - \frac{2}{n-o} + \frac{4}{3} \frac{1}{(n-o)^2}\Big) + a^2\Big(\frac{2o}{(n-o)n} - \frac{8o}{3(n-o)^2n} + \frac{4}{3(n-o)^2}\Big).
\end{split}
\label{eq:4}
\end{equation}

Note that $\mathbbm{E}[S_o] =   \mathbbm{E}[V_1] + ... + \mathbbm{E}[V_o]= \frac{ao}{n}$.


We next relate $V_o$, the amount received by the $o$th recipient, to $S_o$:

\begin{equation*}
\begin{split}
\text{Var}(V_o) & = \mathbbm{E}\Big[\frac{1}{12} \times \frac{(2(a-S_(o-1)))^2}{(n-o+1)^2}\Big] = \frac{1}{3} \mathbbm{E} \Big[\frac{(a-S_{o-1})^2}{(n-o+1)^2} \Big];\\
\text{Var}(V_{o-1}) & = \mathbbm{E}\Big[\frac{1}{12} \times \frac{(2(a-S_(o-2)))^2}{(n-o+2)^2}\Big] = \frac{1}{3} \mathbbm{E} \Big[\frac{(a-S_{o-2})^2}{(n-o+2)^2} \Big].
\end{split}
\end{equation*}

Dividing the first equation by the second, we obtain
$$
\frac{\text{Var}(V_o)}{\text{Var}(V_{o-1})} = \frac{(n-o+2)^2}{(n-o+1)^2} \times \frac{\mathbbm{E}[(a-S_{o-1})^2]}{\mathbbm{E}[(a-S_{o-2})^2]}.
$$

Then,
\begin{equation*}
\begin{split}
\frac{\text{Var}(V_o)}{\text{Var}(V_{1})}  = &
\frac{\text{Var}(V_o)}{\text{Var}(V_{o-1})} \times \frac{\text{Var}(V_{o-1})}{\text{Var}(V_{o-2})} \times ... \times \frac{\text{Var}(V_2)}{\text{Var}(V_{1})} = \\
= & \frac{(n-o+2)^2}{(n-o+1)^2} \times \frac{(n-o+3)^2}{(n-o+2)^2} ...  \times \frac{n^2}{(n-1)^2} \\  
& \times  \frac{\mathbbm{E}[(a-S_{o-1})^2]}{\mathbbm{E}[(a-S_{o-2})^2]} \times  \frac{\mathbbm{E}[(a-S_{o-2})^2]}{\mathbbm{E}[(a-S_{o-3})^2]} \times ... \times \frac{\mathbbm{E}[(a-S_{1})^2]}{a^2} \\ 
= & \frac{n^2}{(n-o+1)^2} \frac{\mathbbm{E}[(a-S_{o-1})^2]}{a^2}.
\end{split}
\end{equation*}

Since $\text{Var}[V_1] = \frac{a^2}{3n^2}$, and $\mathbbm{E}[S_{o-1}] = \frac{a(o-1)}{n}$, we obtain

\begin{equation}
{\text{Var}(V_o)}  = \frac{a^2}{3(n-o+1)^2} \Big( 1- \frac{2(o-1)}{n} \Big) + \frac{1}{3(n-o+1)^2} \mathbbm{E}[S_{o-1}^2].
\label{eq:5}
\end{equation}

Combining Equations~\ref{eq:4} and \ref{eq:5}, we have 

\begin{equation*}
\text{Var}(V_{o+1})  = \Big(1 + \frac{1}{3(n-o)^2}\Big) \text{Var}(V_o).
\end{equation*}

Therefore, we know $1<o<n$:
\begin{equation}
\text{Var}(V_{o})  =  \text{Var}(V_{1}) \prod_{k=1}^{o-1} \Big(1 + \frac{1}{3(n-k)^2}\Big) = \frac{a^2}{3n^2} \prod_{k=1}^{o-1} \Big(1 + \frac{1}{3(n-k)^2}\Big).
\label{eq:6}
\end{equation}

For $o=n$, because the last two recipients split the surplus uniformly at random,  $V_{n-1}$ and $V_{n}$ are identically distributed. Therefore,

$$\text{Var}(V_{n}) = \text{Var}(V_{n-1}) = \frac{a^2}{3n^2} \prod_{k=1}^{n-2} \Big(1 + \frac{1}{3(n-k)^2}\Big).$$
In summary,

\begin{equation}
\text{Var}(V_{o}) =    \begin{cases}
   0 & n=1\text{ and }o = n \\
     \frac{a^2}{3n^2} \prod_{k=1}^{o-1} \Big(1 + \frac{1}{3(n-k)^2}\Big) & n > 1\text{ and } o < n \\
      \frac{a^2}{3n^2} \prod_{k=1}^{n-2} \Big(1 + \frac{1}{3(n-k)^2}\Big) & n > 1 \text{ and } o = n
    \end{cases}
\end{equation}

Furthermore, the variance increases with $o$ when $o<n$.

\section{Additional Analyses}

\subsection{Sample selection}\label{sec:si:sample}

\subsubsection*{Threshold selection.}



First, we illustrate the selection process of a spontaneous red packet (circled in Figure~\ref{fig:cutoff}). If the interval of two consecutive red packets is greater than $\tau$, we divide these two red packets into two ``sessions.'' Therefore, in each session, the time interval between any two consecutive red packets is less than $\tau$. 

We use 24 hours for our timeframe in the main analyses because a red packet expires  24 hours after being sent. To examine the sensitivity of our results to the selection of a 24-hour window, we also select 6, 12, and 48 hours and re-run our analyses; their respective treatment effects are shown in Figure~\ref{fig:threshold}. We find similar results for a 48-hour time window, and a slightly smaller treatment effect for a 6- or 12-hour time window. 


\subsubsection*{Gambling and unnamed groups.}



To explore the impact of our filtering process on our results, we re-run our regressions focusing on gambling groups, for which group names indicate red packet games or gambling, and unnamed groups with functions that are unclear. The results in Figure~\ref{fig:gambling} show that the filtered groups appear to have a higher marginal effect. These results suggest that filtering out these gambling groups may have helped us obtain a more accurate understanding of gift contagion. 

\subsection{Additional results}
\subsubsection*{Alternatives for the econometric model.}\label{sec:si:econometric}


If we apply a linear specification for the effect of the three-dimensional vector --- the total amount of the red packet ($A_r$), the number of recipients ($N_r$), and the order of receiving time ($O_{ir}$), we obtain the following regression:

\begin{equation}
Y_{gir} = \beta T_{gir} + \gamma_1 A_r + \gamma_2 N_r + \gamma_3 O_{ir} + \epsilon_{gir}.
\label{eq:linear}
\end{equation}


As shown in Figure~\ref{fig:si:linear}, the regression results suggest a much larger marginal effect than the results in the main text. One possibility is that the effect of $(A_r, N_r, O_{ir})$ on $Y_{gir}$ is not a linear combination of the three variables, which raises the issue of functional form misspecification. 

\subsubsection*{Direct and indirect reciprocity.}
\label{sec:direct}
Here we present an additional analysis to show that, compared to direct reciprocity, indirect reciprocity plays a dominant role in promoting gift contagion. 
We separate $Y_{gir}$ in into two components: $Y_{gir}^{(1)}$ and $Y_{gir}^{(2)}$. $Y_{gir}^{(1)}$ is the amount sent to the original sender (the sender of red packet $r$), which measures  \textit{direct reciprocity}. $Y_{gir}^{(2)}$ is the amount sent to other group members, which could be a proxy for \textit{indirect reciprocity}. \footnote{Note that a small proportion of the total amount does not belong to either $Y_{gir}^{(1)}$ or $Y_{gir}^{(2)}$ because the sender can also receive a share of her own red packet.} {As shown in Figure~\ref{fig:direct}, on average, the marginal effect on the amount received by the original sender is 3.1\% ($p<0.01$) in the next 24 hours. By contrast, this effect size is much larger for the amount received by other group members: 10.4\% ($p<0.01$). }

\subsubsection*{``Luckiest draw'' and fairness concerns.}

\label{sec:ine}

To investigate whether the fairness concern plays a role in affecting the amount that a user sends, we run the following regression for luckiest draw recipients:

\begin{equation}
Y_{gir} = \beta T_{gir} +
 \alpha_{\operatorname{ratio}} Z_{gir} + \sum_s \gamma_s B_s({A_{r}, N_{r}, O_{ir}})+ \epsilon_{gir}.
\label{eq:inequal}
\end{equation}

\noindent Here we include an additional independent variable: $Z_{gir}$. Let $T_{gir}'$ is the second-largest amount received from the same red packet; and then $Z_{gir} = \frac{T_{gir}'}{T_{gir}}$ represents the ratio of the second-largest amount to the largest amount. We remove the observations (luckiest recipients) that do not have corresponding the second luckiest recipient. Table~\ref{tab:second_largest} reports the regression results.

We find that the ratio of the second-largest to the largest amount  has a negative impact on the likelihood of sending red packets. For example, for the next 10 minutes and 24 hours, we have $\hat{\alpha}_{\operatorname{ratio}}=-0.0363$ and $-0.0379$ for extensive margin ($p<0.01$). 
This suggests that, when the cash amount received by the luckiest-draw recipient is much larger than that received by others, the recipient may feel more obligated to send red packets to the group because of her fairness concern.

\subsubsection*{Age and gender effects.}

\label{sec:si:hetero}
\label{sec:si:additional}

Table~\ref{tab:age} presents our results for different age groups. 
When decomposing the sample by recipients' age, we find that users under 20 years old have the smallest overall and extensive margin compared to those in other age groups. Users over 40 years old have the largest overall and extensive margin, probably because of their higher socioeconomic status. 
When decomposing the sample by senders' age, we find a decreasing trend as the sender's age grows. This result shows that older recipients tend to be responsive to a red packet sent by younger people (Table~\ref{tab:age_sender}). 

We also examine gender differences. 
We do not find any significant gender differences when running regressions on female and male recipients separately (Table~\ref{tab:gender}). 
As shown in Table~\ref{tab:sender_gender}, although we do not find a significant difference in the overall effect between female senders and male senders, 
we find that the red packets sent by female senders exhibit {a significantly higher extensive margin than those sent by male senders ($p<0.01$ for both 10 minutes and 24 hours). }



\subsubsection*{Effect of overall clustering.}
\label{sec:overall_clustering}

Here we use overall clustering as an alternative to the average normalized degree, as a measure of group network structure. The overall clustering of group $g$ is define as 

\begin{equation}
\text{overall clustering}(g) = \frac{ \sum_{i \in |\mathcal{G}| }  \#\{(j, k)| j,k \in  \mathcal{N}_i^g \text{ and } i \neq j \neq k \text{ and } k \in \mathcal{N}_j^g\} }{ \sum_{i \in |\mathcal{G}| }  \#\{(j, k)| j,k \in  \mathcal{N}_i^g \text{ and } i \neq j \neq k\}}.
\end{equation}

\noindent $\mathcal{G}$ denotes the set of members of group $g$, and $\mathcal{N}_i^g$ denotes the set of network neighbors of user $i$ in group $g$.

We report the regression results in Table~\ref{tab:avg_clustering}. 
As shown in Columns (1) and (2), a larger overall clustering predicts a larger amount sent within a group. However, the interaction terms are negative ($p>0.1$ and $p=0.029$ for 10 minutes and 24 hours, respectively),
suggesting that groups with a larger overall clustering do not necessarily induce stronger gift contagion. 
In Columns (3) and (4), the interaction terms are significant and negative, suggesting that for groups with a larger overall clustering, there is generally a smaller extensive margin. For the intensive margin (Columns (5) and (6)), we do not find any significance in the interaction terms. 

\section{Theoretical Framework}

\label{sec:theory_framework}

We form a simple model to motivate our hypotheses by adapting the $n$-player model in \cite{charness2002understanding}. 
Specifically, we use a three-player model, since three is the smallest number by which we can capture  social contagion in groups. 
Without loss of generality, we analyze the sending behavior of player 1 and her utility function is below: 

\begin{equation}
\begin{split}
U_1 (\pi_1, \pi_2, \pi_3) = (1-\lambda) \pi_1 + \lambda \Bigg( \delta \min \left \{\pi_1, \pi_2  - \theta m_2, \pi_3  - \theta m_3 \right \} + \\
(1 - \delta) \left( \pi_1 + ( 1 + \phi m_2 ) \pi_2 + ( 1 + \phi m_3 ) \pi_3 \right)  \Bigg).
\end{split}
\end{equation}

Here we denote the payoffs of the three group members by $\pi_1$, $\pi_2$, and $\pi_3$, respectively. The agent's  weight on their own utility versus social welfare is captured by $\lambda \in [0, 1]$.
The social welfare has two components: a Rawlsian welfare function that cares
about the worst off member (with a weight of $\delta$) and a utilitarian component that cares about the sum of payoffs (with a weight of $1 - \delta$).
$m_2 \geq 0$ and $m_3 \geq 0$, which are two non-negative parameters, denote the merit that player 1 assigns to players 2 and 3, respectively. 
This merit is considered in both Rawlsian welfare component and utilitarian welfare component. 
To account for reciprocity, we introduce two non-negative factors, $\theta$ and $\phi$, in the Rawlsian welfare component and utilitarian welfare component, respectively. 
Depending on the Rawlsian welfare component, the merit term would make a player more deserving as $\theta$ increases; 
depending on the utilitarian welfare component, the merit term would make a player more deserving as $\phi$ increases.

Utilizing the fact that $\pi_1 + \pi_2 + \pi_3 = 0$, we obtain 
\begin{equation}
U_1 (\pi_1, \pi_2, \pi_3) = 
(1-\lambda) \pi_1 + 
\lambda \Bigg( \delta \min \left \{\pi_1, \pi_2  - \theta m_2, \pi_3  - \theta m_3 \right\} + (1 - \delta) \phi \left(  m_2  \pi_2 +  m_3 \pi_3 \right)  \Bigg).
\end{equation}

We then analyze player 1's sending decision as follows. 
Without loss of generality, we assume that player 3 first sends a red packet with the amount of $X=x_1+x_2$ where $x_1$ and $x_2$ are the cash amounts received by players 1 and 2 respectively.
The decision variable is the cash amount that player 1 would send afterwards, which is denoted by $Y$. 
$Y \in [0, \bar{Y}]$ where $\bar{Y}$ is player 1's budget constraint. 
Moreover, we simplify our analysis by considering the expected amount received by players 2 and 3 each, i.e., both players 2 and 3 receive $\frac{Y}{2}$.\footnote{The rationale of the simplification is as follows.
First, our analysis assumes that senders do not receive their own red packets, since we can treat the portions received by their own as if they were not sent out. Then, we can easily derive that the expected amount received by both player 2 and 3 is Y/2: if the number of recipients specified by the sender is 2, according to the gift assignment algorithm, the expected amount received is $Y/2$ for each player; if the number of recipients specified by the sender is 1, the expected amount received is still $Y/2$ once players 2 and 3 have the same probability to be the first one to open the red packet.}
Additionally, we assume that $m_j$ is proportional to the amount received from $j$; then we have $m_2= 0$ and $m_3=\alpha x_1$  where $\alpha > 0$.

Ultimately, we have:\footnote{Since $x_2 + \frac{Y}{2} > - X + \frac{Y}{2} - \theta \alpha x_1$, we omit $x_2 + \frac{Y}{2}$ in the $\min$ term.}
\begin{equation}
\begin{split}
    U_1 & (x_1 - Y, x_2 + \frac{Y}{2}, -X + \frac{Y}{2})   = (1 - \lambda) (x_1 - Y) + \\ & \lambda \left( \delta \min \{ x_1 - Y,  -X + \frac{Y}{2} - \theta  \alpha x_1 \} + (1 - \delta) \phi (  -X + \frac{Y}{2}) \alpha x_1 \right).
\end{split}
\label{eq:u1}
\end{equation}

Therefore, we can rewrite the utility function of player 1 as follows.






\begin{equation}
U_{1}(Y)  =  \begin{cases} 
      (1 - \lambda) (x_1 - Y) +  \lambda \left( \delta (-X + \frac{Y}{2} - \theta  \alpha x_1 )  + (1 - \delta) \phi ( - X + \frac{Y}{2}) \alpha x_1 \right)  &  \text{ if } Y \leq \frac{2}{3}(  ( 2 + \theta \alpha ) x_1 + x_2);\\
      (1 - \lambda) (x_1 - Y) +  \lambda \left( \delta (x_1 - Y ) + (1 - \delta) \phi   ( - X + \frac{Y}{2}) \alpha x_1  \right) &\text{ if } Y > \frac{2}{3}( ( 2 + \theta \alpha ) x_1 + x_2 ).
  \end{cases}
\end{equation}

The derivative of $U_{1}(Y)$ is:
\begin{equation}
 U_{1}'(Y)  =  \begin{cases} 
    -(1-\lambda) + \lambda \delta/2 + (1 - \delta) \phi  \alpha x_1/2, &\text{ if } Y <  \frac{2}{3}( ( 2 + \theta \alpha ) x_1 + x_2 ); \\
    -(1-\lambda) - \lambda \delta  + (1 - \delta) \phi \alpha x_1/2, & \text{ if } Y >  \frac{2}{3}( ( 2 + \theta \alpha ) x_1 + x_2 ).
  \end{cases}
\end{equation}

\noindent We define $u_1' = -(1-\lambda) + \lambda \delta/2 + (1 - \delta) \phi  \alpha x_1/2$, $u_2' = -(1-\lambda) - \lambda \delta  + (1 - \delta) \phi  \alpha x_1/2$.
It is easy to show that $u_1' > u_2'$. Let $\underline{\lambda} = \frac{2 - (1-\delta) \phi \alpha x_1}{2 + \delta}$ and $\bar{\lambda} = \frac{2 - (1-\delta) \phi \alpha x_1}{2 - 2 \delta}$; then the optimal sending amount for player 1: $Y^*$ is below: 
%
%
\begin{equation}
Y^* = \begin{cases}
 \frac{2}{3}( ( 2 + \theta \alpha ) x_1 + x_2 ), & \underline{\lambda} \leq \lambda < \bar{\lambda}; \\
0, & \lambda < \underline{\lambda}; \\
\bar{Y}, & \lambda \geq \bar{\lambda}.
  \end{cases} 
  \label{eq:Y*}
\end{equation}

The strength of the gift contagion is captured by the partial derivative $\frac{\partial Y^*}{\partial x_1}$.
Since $Y^*$ is non-decreasing in $x_1$, we should observe that a player who receives more would send more as well. This leads to our first claim. 

\noindent \textbf{Remark 1} (\textit{Gift contagion}). $\frac{\partial Y^*}{\partial x_1} \geq 0$.

Furthermore, as $Y^*$ is a function of $\theta$, we analyze how the strength of gift contagion varies with $\theta$. We assume  that $\theta$ is  a linear combination of user-, group-, and time parameters, as specified below.
\begin{equation}
\theta = \theta_0 + \theta_{\text{fest}}  \mathbbm{1}_{\text{fest}} + \theta_{\text{group}} + \theta_{\text{luck}}  \mathbbm{1}_{\text{luck}} + \theta_{\text{relation}}.
\end{equation}




First, we consider how the strength of gift contagion varies between festival and non-festival seasons. People expect more frequent gift exchanges during festival periods, such as the exchange of red packets during the Lunar New Year in both East and Southeast Asia and physical gifts in the Western cultures \citep{siu2001red,wang2008significance}. 
We believe that when a red packet is sent during a festival, reciprocity would be more of a consideration, which is captured by 
$\theta_{\text{fest}} \in [0, 1)$.
$\mathbbm{1}_{\text{fest}}$ is an indicator function of whether the red packet is received during a festival period. 
Indeed, we observe that when a red packet is sent during a festival, the gift contagion is stronger.
%

\noindent \textbf{Remark 2} (\textit{Festival effect}). 
$\frac{\partial Y^*}{\partial x_1}  (\mathbbm{1}_\text{fest}=1) \geq \frac{\partial Y^*}{\partial x_1} (\mathbbm{1}_\text{fest}=0)$.


Second, we consider the relationships within groups as measured by $\theta_\text{group}$. 
When the group is one that cares more about reciprocity, we would expect $\theta$, {the concern for reciprocity, is enlarged by a positive $\theta_{\text{group}}$.}

\noindent \textbf{Remark 3} (\textit{Group type effect}).
$\frac{\partial^2 Y^*}{\partial x_1 \partial  \theta_\text{group} } \geq 0 $.

Since red packets are traditionally sent among relatives (see \textit{Appendix~\ref{sec:history}} for more details), we expect $\theta_{\text{group}}$ would be larger in these groups and the partial derivative $\frac{\partial Y^*}{\partial x_1}$ also would increase accordingly. 

Next, we use $\mathbbm{\theta}_{\text{luck}} \in [0, 1)$ to capture the conjecture that the salience of the luckiest draw recipient information may motivate the luckiest draw recipient to send red packets. When the recipient is the luckiest draw, $\mathbbm{1}_{\text{luck}} = 1$; otherwise, $\mathbbm{1}_{\text{luck}} = 0$. Since Eq.~(\ref{eq:Y*}) implies that $Y^{*}$ is non-decreasing in $\theta$, we make the following claim:

\noindent \textbf{Remark 4} (\textit{Luckiest draw effect}).
$\frac{\partial Y^*}{\partial x_1} (\mathbbm{1}_\text{luck}=1) \geq \frac{\partial Y^*}{\partial x_1}(\mathbbm{1}_\text{luck}=0)$.

Finally, we consider how the relationships between player 1 and the other group members affects player 1's  giving, which is captured by 
$\theta_{\text{relation}} \in [0, 1]$. 

\noindent \textbf{Remark 5} (\textit{Effect of individual network position on gift contagion}).
$\frac{\partial^2 Y^*}{\partial x_1 \partial  \theta_\text{relation} } \geq 0 $.

 Prior research  has shown that social contagion is stronger for people who  have a larger network degrees \citep{kempe2003maximizing,centola2007complex,bakshy2012role,yuan2021causal} or who are less clustered in networks \citep{aral2012identifying,ugander2012structural}. 
 We thus expect $\theta_{\text{relation}}$ to be larger for those people with more within-group online friends (i.e., degree), as well as for those who are less clustered in their within-group social network.

\clearpage




\renewcommand{\thetable}{\Roman{table}}

\clearpage

\setcounter{table}{0}    

\renewcommand\thefigure{A.\arabic{figure}}    
\renewcommand\thetable{A.\arabic{table}}    

\begin{table*}[h!]
	\centering
\linespread{1.25}\selectfont
	\caption{Summary statistics of online groups}
	\label{tab:group}
	\small
		\begin{tabular}{lcccccc}
			\hline \hline
			Variables       & Mean   & Min  & 25\%   & 50\%   & 75\%    & Max       \\ \hline
			Group size             & 19.91  & 3   &    8.0 &   14.0  & 24.0 &   490       \\
			Total no. of red packets      & 210.24 & 9  &    54.0 & 115.0  & 253.5 & 8458\\
			  Total cash amount of red packets (CNY)    & 919.30 & 0.11  & 164.46 & 418.86 &  990.0 & 373679.07     \\
			\hline \hline
		\end{tabular}
\end{table*}


\begin{table}[h!]
\centering
\linespread{1.25}\selectfont
\caption{Summary statistics of sample users (1)}
\small
\label{tab:user1}
\begin{tabular}{lcc}
\hline\hline
\textbf{Variables} & \textbf{Count}     & \textbf{Proportion} \\ \hline
\textit{Gender}               &           &            \\
Male                 & 1,783,737 &51.69\%    \\
Female               & 1,639,955 &47.53\%    \\
Unreported           & 26,848 & 0.78\%\\
\textit{Age}                  &           &            \\
{[}10, 20)           &  364,352 & 10.56\% \\
{[}20, 30)           &  1,724,020 & 49.96\% \\
{[}30, 40)           &  943,051 & 27.33\%   \\
{[}40, 50)           &  274,443 &  7.95\%    \\
{[}50, 60)           & 39,522 &    1.15\% \\
Other/Unreported     & 105,152 &   3.05\%   \\ \hline\hline
\end{tabular}
\end{table}


\begin{table*}[h!]
	\centering
	\caption{Summary statistics of  sample users (2)}
\linespread{1.25}\selectfont
	\small
	\label{tab:user2}
	\begin{threeparttable}
	\linespread{1.25}\selectfont
		\begin{tabular}{lcccccc}
			\hline \hline
			\textbf{Variables}       & \textbf{Mean}   & \textbf{Min}  &\textbf{ 25\%}   & \textbf{50\% }  &\textbf{ 75\% }   & \textbf{Max}       \\ \hline
			Within-group degree   & 8.75 & 1 & 4 & 7 & 11 & 358            \\
			  No. of private contacts    & 182.61 & 0 & 54 & 110 & 204 & 25,956     \\
			  No. of groups that a user joins    & 38.99 & 1 & 9 & 20 & 40 & 16,945,750     \\
			\hline \hline
			\end{tabular}
	\end{threeparttable}
\end{table*}


\begin{table*}[h!]
\centering
\linespread{1.25}\selectfont
\caption{Summary statistics of red packets}
\small
\label{tab:rp}
\makebox[\linewidth]{
\begin{threeparttable}
\begin{tabular}{lcccccc}
\hline\hline
\textbf{Statistics}       & \textbf{Mean}   & \textbf{Min}  & \textbf{25\%}   & \textbf{50\%}   & \textbf{75\% }   & \textbf{Max}       \\ \hline
Amount              & 4.37 & 0.01 & 0.5 & 1 & 5 & 200      \\
No. of recipients   &  5.06  & 1    & 3   & 5    & 5   & 100         \\
Time interval between red packets {(in seconds)}\tnote{a}\  & 29304.07 & 0 & 46 & 130 & 938&  12,475,671 \\
Completion time (in seconds)\tnote{b} & 1267.53 & 2 & 10 &  23 & 176 & 509,131\\ \hline\hline
\end{tabular}
\begin{tablenotes}
\small{
\item[a] Time interval indicates the time between the current and the preceding red packet. 
\item[b] Completion time measures the time interval between the red packet's sending time and the time when the last share of this red packet was received. Red packets that are not received by anyone are excluded.}
\end{tablenotes}
\end{threeparttable}
}
\end{table*}

\clearpage

\begin{table*}[h!]
\centering
\linespread{1.25}\selectfont
\caption{Regression coefficients and the corresponding adjusted $p$ values for red packets with the amount of 5 CNY and 3 recipients. $(a, n, o)$ refers to the total amount, the number of recipients, and the order of receiving time.}
\footnotesize
\label{tab:prandom1}
\makebox[\linewidth]{
\begin{tabular}{lccccccccccccc}
\hline \hline 
Variables	&	(5,3,1)	& Adj.	$p$	&	(5,3,2)	& Adj.	$p$	&	(5,3,3)	&	Adj. $p$	&	\\ \hline 
female	&	0.0016	&	0.9365	&	0.0049	&	0.8036	&	0.0019	&	0.9296	&	\\
age	&	-0.0029	&	0.9458	&	0.0379	&	0.9023	&	-0.0640	&	0.9023	&	\\
degree	&	0.0520	&	0.9023	&	-0.0806	&	0.9023	&	0.0090	&	0.9365	&	\\
fricnt	&	0.5963	&	0.9365	&	0.1410	&	0.9365	&	2.4340	&	0.8036	&	\\
joincnt	&	0.3196	&	0.9023	&	0.1584	&	0.9365	&	0.3770	&	0.9023	&	\\
history\_sendamt	&	-0.6952	&	0.9365	&	0.1280	&	0.9365	&	0.1321	&	0.9365	&	\\
history\_sendcnt	&	-0.2336	&	0.9023	&	-0.0903	&	0.9365	&	0.1838	&	0.9023	&	\\
history\_recvamt	&	-0.9458	&	0.9023	&	0.1200	&	0.9365	&	0.2068	&	0.9365	&	\\
history\_recvcnt	&	-0.5519	&	0.9023	&	-0.0977	&	0.9365	&	0.4581	&	0.9023	&	\\
groupamt	&	-3.6014	&	0.9365	&	-1.1416	&	0.9365	&	0.9528	&	0.9365	&	\\
groupnum	&	-7.6666	&	0.9023	&	-5.3852	&	0.9296	&	3.9070	&	0.9365	&	\\
 \hline \hline
\end{tabular}}
\end{table*}


\begin{table*}[h!]
\centering
\footnotesize
\linespread{1.25}\selectfont
\caption{Regression coefficients and the corresponding adjusted $p$ values for red packets with the amount of 10 CNY and 5 recipients. $(a, n, o)$ refers to the total amount, the number of recipients, and the order of receiving time.}
\setlength{\tabcolsep}{3pt}
\label{tab:prandom2}
     \makebox[\linewidth]{
     \linespread{1.25}\selectfont
\begin{tabular}{lcccccccccccccccc}
\hline \hline
Variable   & (10,5,1) & Adj. $p$ & (10,5,2) & Adj. $p$ & (10,5,3) &  Adj. $p$ & (10,5,4) & Adj. $p$ & (10,5,5) & Adj. $p$ &\\ \hline
female	&	0.0006	&	0.9490	&	-0.0002	&	0.9490	&	-0.0007	&	0.9490	&	-0.0007	&	0.9490	&	0.0022	&	0.6569	\\
age	&	0.005	&	0.9490	&	0.0138	&	0.9490	&	0.0161	&	0.9490	&	-0.0341	&	0.7052	&	-0.0097	&	0.9490	\\
degree	&	0.0922	&	0.8288	&	-0.0785	&	0.6569	&	0.003	&	0.9556	&	0.0048	&	0.9490	&	0.0112	&	0.9490	\\ 
fricnt	&	-1.5021	&	0.6569	&	1.4011	&	0.6569	&	-1.1583	&	0.7047	&	-0.6723	&	0.8807	&	-0.4135	&	0.9490	\\
joincnt	&	-0.0285	&	0.9490	&	0.0302	&	0.9490	&	-0.0339	&	0.9490	&	-0.2852	&	0.6569	&	-0.095	&	0.9490	\\
history\_sendamt	&	1.0972	&	0.8132	&	0.8839	&	0.8288	&	2.1079	&	0.6569	&	-0.2787	&	0.9490	&	-0.32	&	0.9490	\\
history\_sendcnt	&	0.0613	&	0.9490	&	0.2013	&	0.6569	&	0.1389	&	0.6569	&	-0.0471	&	0.9490	&	-0.0268	&	0.9490	\\
history\_recvamt	&	1.3355	&	0.6569	&	0.1375	&	0.9490	&	1.4951	&	0.6569	&	-0.5232	&	0.9119	&	0.0097	&	0.9875	\\ 
history\_recvcnt	&	0.6882	&	0.6569	&	0.478	&	0.7167	&	0.7242	&	0.6569	&	-0.2521	&	0.9119	&	-0.3717	&	0.7052	\\
groupamt	&	29.2084	&	0.6569	&	4.5624	&	0.9490	&	23.1333	&	0.6569	&	-1.8278	&	0.9490	&	3.9539	&	0.9490	\\
groupnum	&	7.2803	&	0.9490	&	13.0648	&	0.7052	&	3.8353	&	0.9490	&	-6.4186	&	0.9358	&	-7.3134	&	0.8807	\\
\hline \hline
\end{tabular}}
\end{table*}


\begin{table}[]
\caption{Regression results for generalized reciprocity }
\scriptsize
\centering
\label{tab:outcontagion}\makebox[\linewidth]{
\begin{threeparttable}
\linespread{1.25}\selectfont
\begin{tabular}{p{4.cm}cccccccccccc}
\hline\hline
& \multicolumn{2}{c}{Overall} &
\multicolumn{2}{c}{Extensive} & 
\multicolumn{2}{c}{Intensive} \\
& 10 min & 24 h  & 10 min & 24 h  & 10 min & 24 h   \\  
& (1) & (2)  & (3) & (4)  & (5) &  (6)   \\ 
\hline 
 Amount received & -0.0028 & 0.1310 & 0.0000 & -0.0002 & 1.0578 & 8.1754 \\
  &(0.0107) & (0.1288) & (0.0001) & (0.0005) & (4.3879) & (7.4618) \\

Stratum fixed effect & Y & Y & Y & Y & Y & Y \\
No. of observations & 154,312 & 154,312 & 154,312 & 154,312 & 321 & 5182   \\
 Adjusted $R^2$ &  -0.0601 & -0.0086 & -0.0040 & 0.0345 & 0.9143 &  0.2648 \\ 
 \hline\hline
\end{tabular}
\begin{tablenotes}[flushleft]
\small
\item \scriptsize \textit{Note}: \linespread{1.25}\selectfont
The dependent variable (DV) for Columns (1) and (2) is the amount sent within the respective timeframe. It is coded as zero for those who do not send red packets.
The DV in Columns (3) and (4) is the dummy variable for sending red packets. 
The DV in Columns (5) and (6) is the amount conditioning on sending red packets. Marginal effects are reported.
Standard errors clustered at the group and user levels are in parentheses. 
*: $p<0.1$, **: $p<0.05$, ***: $p<0.01$. 
\end{tablenotes}
\end{threeparttable}}
\end{table}


\begin{table}[]
\caption{Regression results for the ratio of the second largest to largest amount received }
\scriptsize
\centering
\label{tab:second_largest}\makebox[\linewidth]{
\begin{threeparttable}
\linespread{1.25}\selectfont
\begin{tabular}{p{4.cm}cccccccccccc}
\hline\hline
& \multicolumn{2}{c}{Overall} &
\multicolumn{2}{c}{Extensive} & 
\multicolumn{2}{c}{Intensive} \\
& 10 min & 24 h  & 10 min & 24 h  & 10 min & 24 h   \\  
& (1) & (2)  & (3) & (4)  & (5) &  (6)   \\ 
\hline 
Amount received & 0.3325*** & 0.3775** & 0.0055*** & 0.0056*** & -0.3949 & -0.9492*  \\
& (0.0918) & (0.1830) & (0.0004) & (0.0005) & (0.3232) & (0.5545) \\
Ratio & 0.0630 & -0.1120 & -0.0363*** & -0.0379*** & 0.2162 & -0.3313 \\
& (0.1142) & (0.2394) & (0.0020) & (0.0022) & (0.4727) & (0.9166)  \\
Stratum fixed effect & Y & Y & Y & Y & Y & Y \\
No. of observations & 1,268,240 &1,268,240 &1,268,240 &1,268,240 &223,329& 273,898   \\
 Adjusted $R^2$ &  0.0620 & 0.0469 & 0.0326 & 0.0350 & 0.1811 &  0.1208 \\ 
 \hline\hline
\end{tabular}
\begin{tablenotes}[flushleft]
\small
\item \scriptsize \mytablenote
\end{tablenotes}
\end{threeparttable}}
\end{table}

\begin{table}[h!]
\caption{Heterogeneous treatment effects for recipient's age}
\scriptsize
\centering
\label{tab:age}\makebox[\linewidth]{
\begin{threeparttable}
\linespread{1.25}\selectfont
\begin{tabular}{p{4.cm}cccccccccccccccccc}
\hline\hline
& \multicolumn{2}{c}{Overall} &
\multicolumn{2}{c}{Extensive} & 
\multicolumn{2}{c}{Intensive} & 
\\
& 10 min & 24 h  & 10 min & 24 h  & 10 min & 24 h\\ 
& (1) & (2)  & (3) & (4)  & (5) & (6) \\  \hline
& \multicolumn{6}{c}{Recipient age 10-19} \\
Amount received & 0.0597* & 0.1063 & 0.0026*** & 0.0028*** & -0.0507 & 0.0914\\
 & (0.0434) & (0.1027) & (0.0004) & (0.0004) & (0.2304) & (0.5112) \\
Stratum fixed effect  & Y  & Y  & Y & Y  & Y  & Y \\
No. of observations & 610,635 & 610,635 & 610,635 & 610,635 & 88,094 & 112,338   \\
Adjusted $R^2$ &   0.1012 & 0.0876 & 0.0097 & 0.0130 & 0.2728 & 0.2117 \\
\hline
& \multicolumn{6}{c}{Recipient age 20-29}\\
Amount received & 0.1904*** & 0.2411*** & 0.0033*** & 0.0033*** & 0.2239** & 0.1560\\
 & (0.0291) & (0.0602) & (0.0001) & (0.0002) & (0.1198) & (0.2175) \\
 Stratum fixed effect  & Y  & Y  & Y & Y  & Y  & Y \\
 No. of observations & 3,635,995 & 3,635,995 & 3,635,995 & 3,635,995 & 500,387 & 648,902 
    \\
Adjusted $R^2$ & 0.0376 & 0.0403 & 0.0194 & 0.0223 & 0.1325 & 0.1055 \\ \hline
& \multicolumn{6}{c}{Recipient age 30-39}\\
Amount received & 0.1354*** & 0.1506* & 0.0031*** & 0.0032*** & -0.2444* & -0.5716**\\
 & (0.0485) & (0.0904) & (0.0002) & (0.0002) & (0.2015) & (0.3209) \\
Stratum fixed effect  & Y  & Y  &  Y  & Y  & Y  & Y   \\ 
No. of observations & 2,169,069 & 2,169,069 & 2,169,069 & 2,169,069 & 323,131 & 418,970 
    \\
Adjusted $R^2$ & 0.0524 & 0.0587 & 0.0176 & 0.0215 & 0.2033 & 0.1595
 \\\hline
& \multicolumn{6}{c}{Recipient age $\geq$40}\\
Amount received & 0.2506*** & 0.3600*** & 0.0033*** & 0.0038*** & 0.1021 & -0.0820\\
 & (0.0579) & (0.1017) & (0.0004) & (0.0004) & (0.2155) & (0.3198) \\
Stratum fixed effect  & Y  & Y  &  Y  & Y  & Y  & Y   \\ 
No. of observations & 633,413 & 633,413 & 633,413 & 633,413 & 111,341 & 142,255 
    \\
Adjusted $R^2$ &  0.0893 & 0.1141 & 0.0085 & 0.0144 & 0.2587 & 0.2345 
 \\\hline\hline
\end{tabular}
\begin{tablenotes}[flushleft]
\small
\item \scriptsize \mytablenote
\end{tablenotes}
\end{threeparttable}}
\end{table}

\begin{table}[h!]
\caption{Heterogeneous treatment effects for sender's age }
\scriptsize
\centering
\label{tab:age_sender}\makebox[\linewidth]{
\begin{threeparttable}
\linespread{1.25}\selectfont
\begin{tabular}{p{4.cm}cccccccccccccccccc}
\hline\hline
& \multicolumn{2}{c}{Overall} &
\multicolumn{2}{c}{Extensive} & 
\multicolumn{2}{c}{Intensive} & 
\\
& 10 min & 24 h  & 10 min & 24 h  & 10 min & 24 h\\ 
& (1) & (2)  & (3) & (4)  & (5) & (6) \\  \hline
& \multicolumn{6}{c}{Sender Age 10-19} \\
Amount received & 0.3197*** & 0.4384*** & 0.0049*** & 0.0046*** & 0.4281 & 0.8123\\
 & (0.0966) & (0.1616) & (0.0009) & (0.0009) & (0.6144) & (0.8911) \\
Stratum fixed effect  & Y  & Y  & Y & Y  & Y  & Y \\
No. of observations & 428,258 & 428,258 &  428,258 &  428,258 &  68,819 &  87,310 \\
Adjusted $R^2$ & 0.2188 & 0.1823 & 0.0128 & 0.0176 & 0.4636 & 0.3754 \\
\hline
& \multicolumn{6}{c}{Sender Age 20-29}\\
Amount received & 0.2041*** & 0.1973*** & 0.0033*** & 0.0034*** & 0.0963 & -0.2579  \\
 & (0.0295) & (0.0626) & (0.0002) & (0.0002) & (0.1330) & (0.2622)  \\
 Stratum fixed effect  & Y  & Y  & Y & Y  & Y  & Y \\
 No. of observations & 3,396,534 & 3,396,534 & 3,396,534 & 3,396,534 &  482,079 & 621,904 \\
Adjusted $R^2$ & 0.0361 & 0.0433 & 0.0252 & 0.0279 &  0.1492 & 0.1344 \\ \hline
& \multicolumn{6}{c}{Sender Age 30-39}\\
Amount received & 0.1266*** & 0.1574*** & 0.0029*** & 0.0030*** & -0.1068 & -0.3465*\\
 &  (0.0285) & (0.0600) & (0.0001) & (0.0001) & (0.1191) & (0.2180)  \\
Stratum fixed effect  & Y  & Y  &  Y  & Y  & Y  & Y   \\ 
No. of observations &  2,475,226 & 2,475,226 & 2,475,226 & 2,475,226 & 345,911 & 449,217 \\
Adjusted $R^2$ & 0.0346 & 0.0323 & 0.0223 & 0.0251 & 0.1471 & 0.1286 \\\hline
& \multicolumn{6}{c}{Sender Age $\geq$40}\\
Amount received & 0.1009*** & 0.1658*** & 0.0029*** & 0.0030*** & -0.1336 & -0.1303 \\
 & (0.0328) & (0.0588) & (0.0003) & (0.0003) & (0.1567) & (0.2214) \\
Stratum fixed effect  & Y  & Y  &  Y  & Y  & Y  & Y   \\ 
No. of observations & 780,907 & 780,907 & 780,907 & 780,907 & 120,486 & 155,465 \\
Adjusted $R^2$ & 0.0595 & 0.0526 & 0.0119 & 0.0153 & 0.2525 & 0.1839 \\\hline\hline
\end{tabular}
\begin{tablenotes}[flushleft]
\small
\item \scriptsize \mytablenote
\end{tablenotes}
\end{threeparttable}}
\end{table}
\clearpage

\begin{table}[]
\caption{Heterogeneous treatment effects for recipient's gender }
\scriptsize
\centering
\label{tab:gender}\makebox[\linewidth]{
\begin{threeparttable}
\linespread{1.25}\selectfont
\begin{tabular}{p{4cm}cccccc}
\hline\hline
& \multicolumn{2}{c}{Overall} &
\multicolumn{2}{c}{Extensive} & 
\multicolumn{2}{c}{Intensive} 
\\
& 10 min & 24 h  & 10 min & 24 h  & 10 min & 24 h   \\ 
& (1) & (2)  & (3) & (4)  & (5) & (6) \\  \hline
& \multicolumn{6}{c}{Female recipient}\\
Amount received & 0.1364*** & 0.1967** & 0.0029*** & 0.0030*** & 0.1540**& 0.0942\\
 & (0.0234) & (0.0438) & (0.0001) & (0.0001) & (0.0897) & (0.1429) \\
Stratum fixed effect  & Y  & Y  &  Y & Y  & Y  & Y  \\
No. of observations & 3,870,582 & 3,870,582 & 3,870,582 & 3,870,582 & 551,408 & 711,086 
    \\
Adjusted $R^2$ &0.0450 & 0.0441 & 0.0188 & 0.0217 & 0.1867 & 0.1444 
 \\  \hline
 & \multicolumn{6}{c}{Male recipient} \\
Amount received & 0.1845*** & 0.1755** & 0.0033*** & 0.0034*** & -0.2148 & -0.6624**\\
 & (0.0415) & (0.0886) & (0.0001) & (0.0002) & (0.1700) & (0.3124) \\
Stratum fixed effect  & Y  & Y  &  Y & Y  & Y  & Y    \\ 
No. of observations & 3,380,557 & 3,380,557 & 3,380,557 & 3,380,557 & 506,889 & 656,573 
    \\
Adjusted $R^2$ &   0.0439 & 0.0454 & 0.0194 & 0.0220 & 0.1587 & 0.1158 \\ 
\hline \hline
\end{tabular}
\begin{tablenotes}[flushleft]
\item \scriptsize \mytablenote
\end{tablenotes}
\end{threeparttable}}
\end{table}


\begin{table}[]
\caption{Heterogeneous treatment effects for sender's gender }
\scriptsize
\centering
\label{tab:sender_gender}\makebox[\linewidth]{
\begin{threeparttable}
\linespread{1.25}\selectfont
\begin{tabular}{p{4cm}cccccc}
\hline\hline
& \multicolumn{2}{c}{Overall} &
\multicolumn{2}{c}{Extensive} & 
\multicolumn{2}{c}{Intensive} 
\\
& 10 min & 24 h  & 10 min & 24 h  & 10 min & 24 h   \\ 
& (1) & (2)  & (3) & (4)  & (5) & (6) \\  \hline
& \multicolumn{6}{c}{Female sender }\\
Amount received & 0.1401*** & 0.1808*** & 0.0036*** & 0.0038*** & -0.1529 & -0.4191* \\
 & (0.0287) & (0.0564) & (0.0002) & (0.0002) & (0.1424) & (0.2654)   \\
Stratum fixed effect  & Y  & Y  &  Y & Y  & Y  & Y  \\
No. of observations & 2,960,098 & 2,960,098 & 2,960,098 & 2,960,098 & 478,056 & 609,017 
    \\
Adjusted $R^2$ & 0.0541 & 0.0471 & 0.0211 & 0.0233 & 0.2318 & 0.1547  \\  \hline
 & \multicolumn{6}{c}{Male sender } \\
Amount received & 0.1571*** & 0.1780*** & 0.0027*** & 0.0028*** & 0.0638 & -0.1679 \\
 & (0.0209) & (0.0434) & (0.0001) & (0.0001) & (0.0916) & (0.1541) \\ 
Stratum fixed effect  & Y  & Y  &  Y & Y  & Y  & Y    \\ 
No. of observations &  4,329,692 &  4,329,692 & 4,329,692 & 4,329,692 & 573,594 & 748,681 \\
Adjusted $R^2$ & 0.0309 & 0.0295 & 0.0247 & 0.0272  & 0.1281 & 0.0995 \\ 
\hline \hline
\end{tabular}
\begin{tablenotes}[flushleft]
\item \scriptsize \mytablenote
\end{tablenotes}
\end{threeparttable}}
\end{table}


\begin{table}[]
    \centering
    \caption{Effect of individual eigenvector centrality on gift contagion }
    \label{tab:eigen}
    \begin{threeparttable}
    \linespread{1.25}\selectfont
    \scriptsize
    \begin{tabular}{lcccccc}
    \hline \hline
    	& \multicolumn{2}{c}{Overall} 	&  	\multicolumn{2}{c}{Extensive}  &  \multicolumn{2}{c}{Intensive}	\\
    	& 10 min	& 24 h	& 10 min &	24 h	& 10 min	& 24 h \\
    	& (1) & (2) & (3) & (4) & (5) & (6) \\  \hline
    Amount received &	0.1651*** &	0.3060** &	0.0060*** &	0.0061*** &	0.1976 &	0.5564 \\
    	& (0.0616) &	(0.1300) &	(0.0002) &	(0.0003) &	(0.2118) &	(0.3815) \\
    Amount received $\times$  eigen &	-0.0279&	-0.3501 &	-0.0086***	& -0.0084*** &	-0.5254 &	-2.1610* \\
    	& (0.1846) &	(0.4085) &	(0.0006) &	(0.0007) &	(0.6127) &	(1.1843) \\
    Eigen &	2.6790*** &	6.1329*** &	0.2158*** &	0.2987*** & 5.5308*** &	9.2871*** \\
    	 & (0.1345) &	(0.2843) &	(0.0023) &	(0.0026) &	(0.6236) &	(1.1001) \\
    Group size & Y & Y & Y & Y & Y & Y \\
    Stratum fixed effect & Y & Y & Y & Y & Y & Y \\ 
    No. of observations	& 7,266,446 &	7,266,446 &	7,266,446	& 7,266,446	& 1,060,746 &	1,370,741	\\
    Adjusted $R^2$ &	0.0398 &	0.0401 & 0.0266	 & 0.0315 &	0.1520 & 0.1099	\\
     \hline  \hline
    \end{tabular}
    \begin{tablenotes}[flushleft]
\item \scriptsize \mytablenote
\end{tablenotes}
\end{threeparttable}
\end{table}


\begin{table}[h]
    \centering
    \caption{Effect of overall clustering in groups  }
        \label{tab:avg_clustering}
    \linespread{1.25}\selectfont
    \scriptsize\makebox[\linewidth]{
    \begin{threeparttable}
    \begin{tabular}{lcccccc}
    \hline \hline
    	& \multicolumn{2}{c}{Overall} 	&  	\multicolumn{2}{c}{Extensive}  &  \multicolumn{2}{c}{Intensive}	\\
    	& 10 min	& 24 h	& 10 min   &	24 h &   10 min &   24 h \\
    	& (1) & (2) & (3) & (4) & (5) & (6) \\  \hline
    Amount received & 0.2864*** & 0.6123*** & 0.0077*** & 0.0080*** & -0.0417 & 0.2352 \\
      & (0.0815) & (0.1876) & (0.0004) & (0.0004) & (0.2587) & (0.4689)  \\
    Amount received $\times$ overall clustering & -0.1718 & -0.5604** & -0.0061*** & -0.0063*** & 0.0621 & -0.5974 \\
     &  (0.1081) & (0.2558) & (0.0005) & (0.0005) & (0.3429) & (0.6558)  \\
    Average overall clustering & 0.7681*** & 1.5241*** & 0.0023 & -0.0006 & 5.5541*** & 8.5444*** \\
     & (0.0914) & (0.2059) & (0.0023) & (0.0028) & (0.4564) & (0.7958) \\
        Group size & Y & Y & Y & Y & Y & Y \\
        Stratum fixed effect & Y & Y & Y & Y & Y & Y \\ 
    No. of observations	& 7,266,446 &	7,266,446 &	7,266,446	& 7,266,446	 & 1,060,746 &	1,370,741\\
    Adjusted $R^2$ & 0.0397 & 0.0399 & 0.0238 & 0.0272 & 0.1523 & 0.1100 \\
     \hline  \hline
    \end{tabular}
    \begin{tablenotes}[flushleft]
     \scriptsize \item \mytablenote
    \end{tablenotes}
    \end{threeparttable}}
\end{table}


\setcounter{figure}{0} 
\renewcommand{\thefigure}{\Roman{figure}}

\clearpage
\setcounter{figure}{0} 
\renewcommand\thefigure{A.\arabic{figure}}

\begin{figure}
    \centering
    \includegraphics[width=\linewidth]{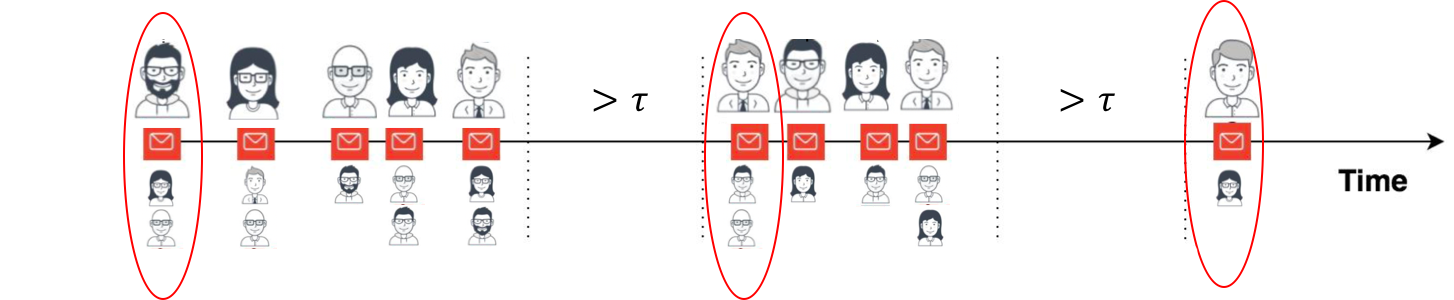}
    \caption{\raggedright \linespread{1.2} \selectfont The illustration of spontaneous red packets (circled red packets) and sessions (three sessions split by dashed lines in this example). }
    \label{fig:cutoff}
\end{figure}


\begin{figure}
    \centering
\includegraphics[width=0.4\linewidth]{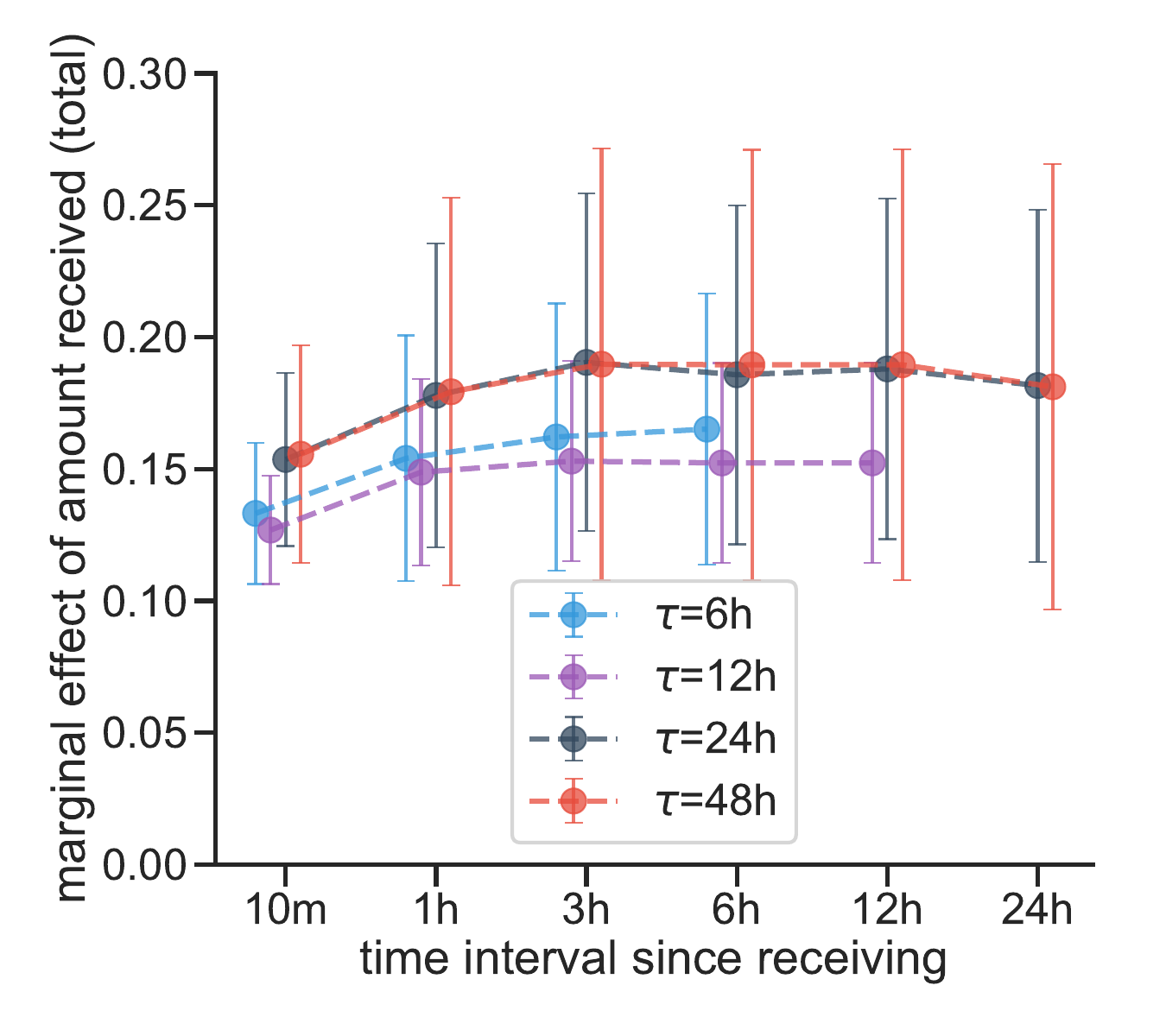}
    \caption{\raggedright \linespread{1.2} \selectfont     Treatment effects for different $\tau$. Error bars are the 95\% CIs clustered at the group- and user-levels.  }
    \label{fig:threshold}
\end{figure}


\begin{figure}
    \centering
    \includegraphics[width=0.4\linewidth]{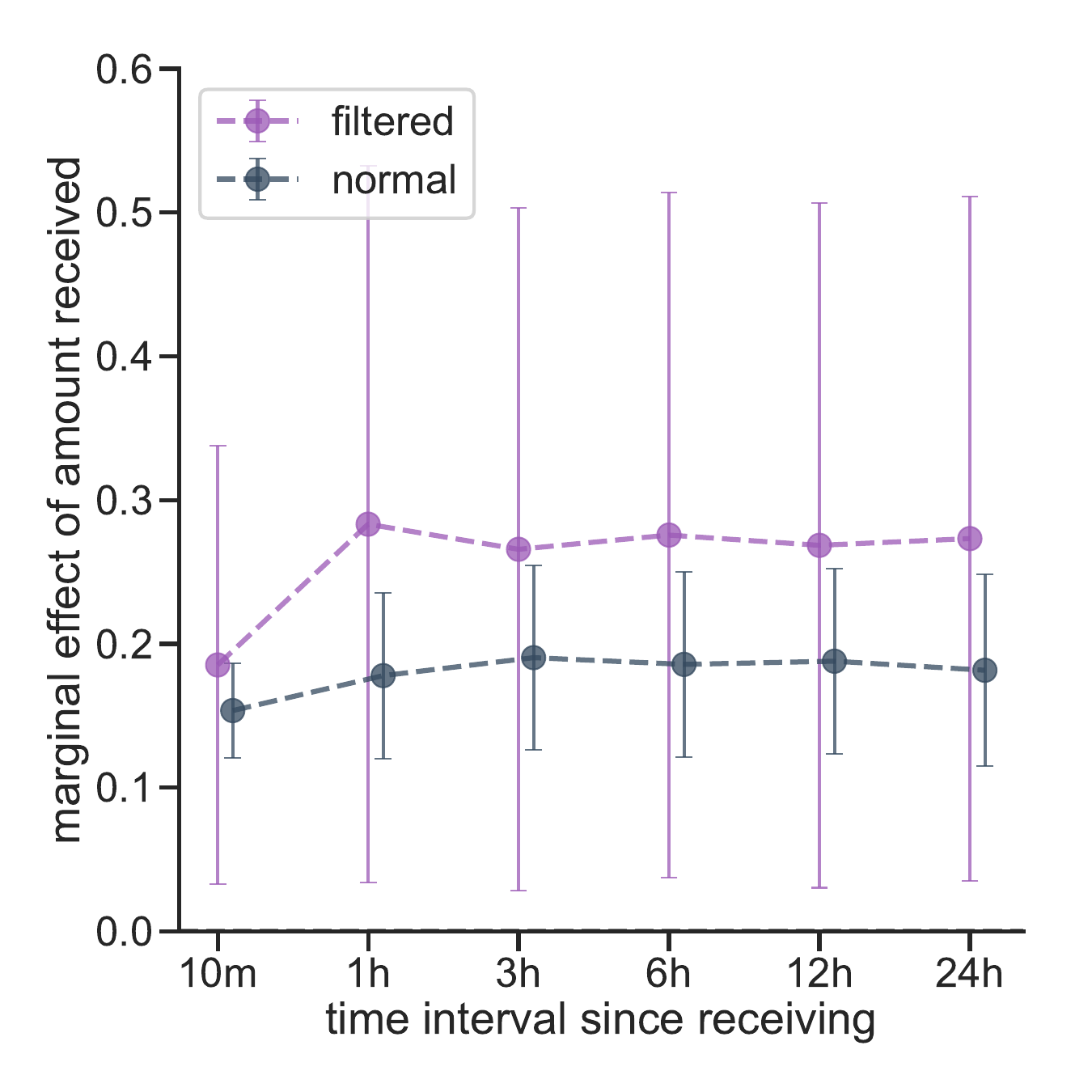}
    \caption{\raggedright \linespread{1.2} \selectfont Treatment effects for normal groups studied in the main text and groups that were filtered out. Error bars are the 95\% CIs clustered at the group- and user-levels.}
    \label{fig:gambling}
\end{figure}

\begin{figure}
    \centering
    \linespread{2}\selectfont
    \includegraphics[width=0.4\linewidth]{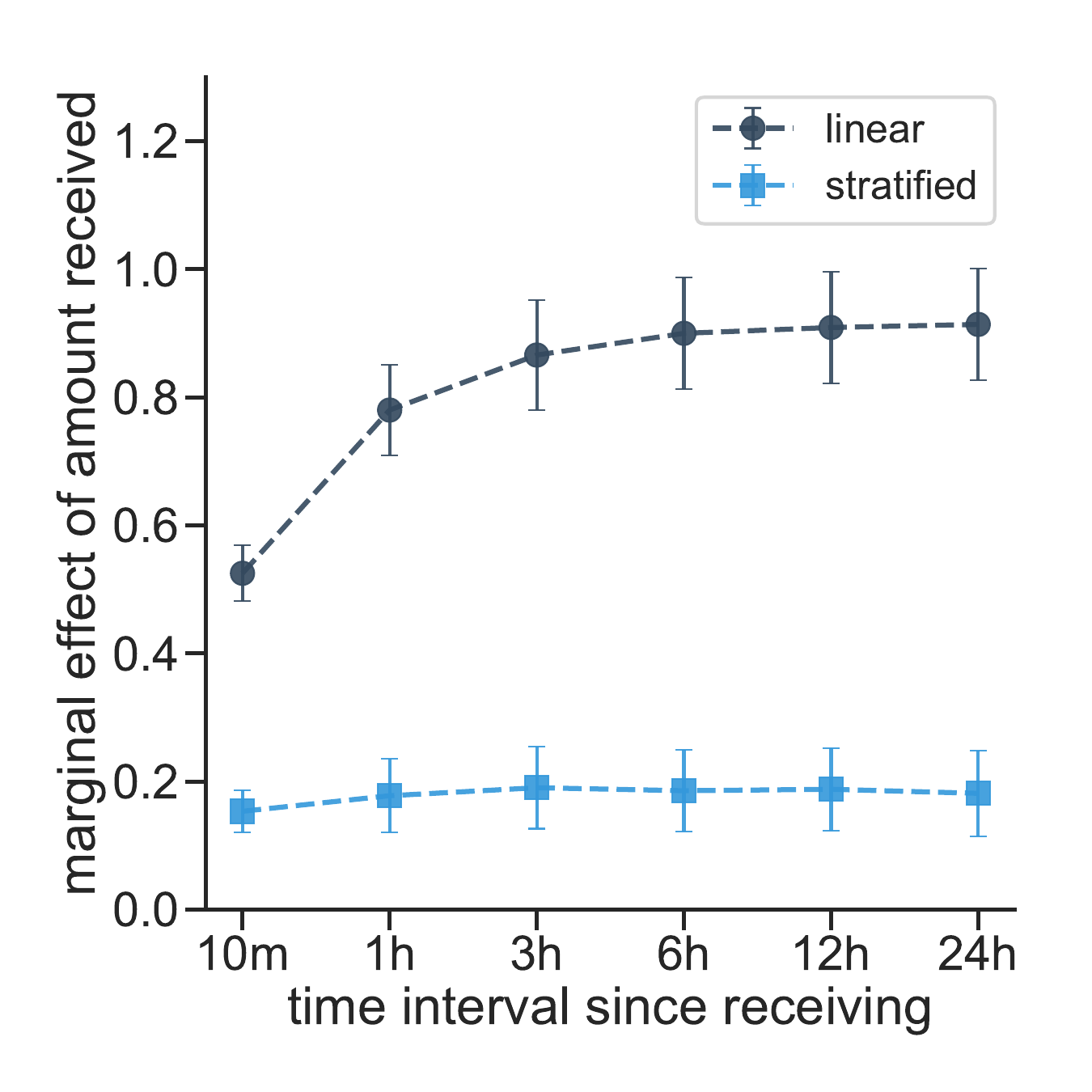}
    \caption{\raggedright \linespread{1.2} \selectfont The regression results for the linear specification of the effect of $(A_r, N_r, O_{ir})$. ``Linear'' represents the results when $(A_r, N_r, O_{ir})$ is linearly specified. ``Stratified'' represents the results when $(A_r, N_r, O_{ir})$ is used to stratify data, as is in the main text. Error bars are the 95\% CIs clustered at the group- and user-levels.}
    \label{fig:si:linear}
\end{figure}


\begin{figure}
    \centering
    \linespread{2}\selectfont
    \includegraphics[width=0.4\linewidth]{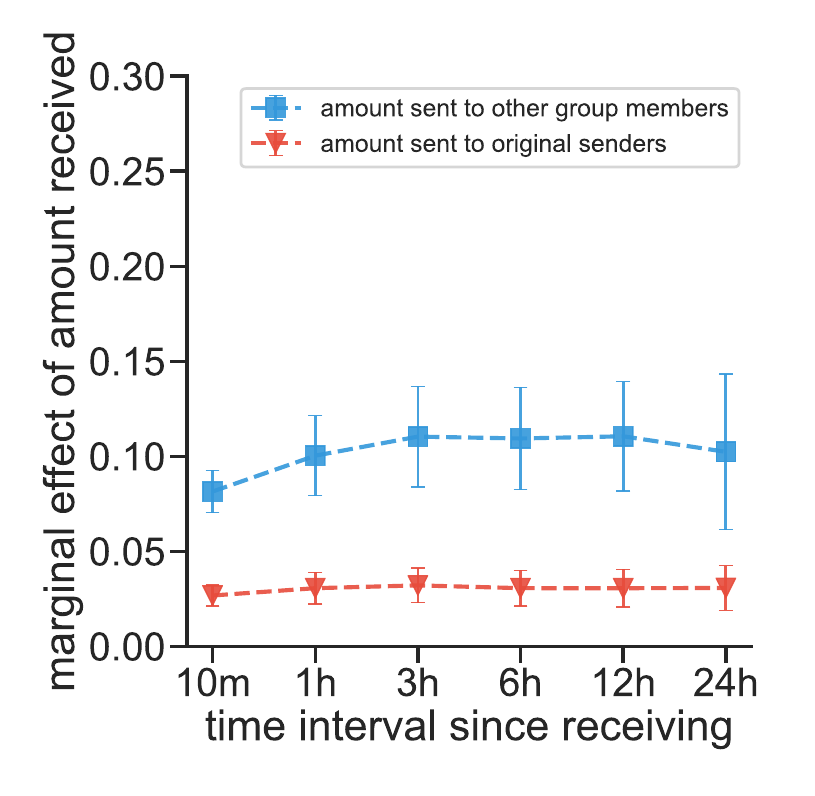}
    \caption{\raggedright \linespread{1.2} \selectfont The $x$-axis represents the time interval since receiving a red packet. The $y$-axis represents the marginal effect of receiving red packets on the amount sent in the future. 
``Indirect'' and ``direct'' refer to the ratio of the amount sent to group members except for the original sender and to the original sender, respectively. 
Error bars are the 95\% CIs clustered at the group- and user-levels.}
    \label{fig:direct}
\end{figure}






\clearpage

\bibliographystyle{informs2014}
\bibliography{main}

\begin{thebibliography}{55}
\providecommand{\natexlab}[1]{#1}
\providecommand{\url}[1]{\texttt{#1}}
\providecommand{\urlprefix}{URL }

\bibitem[{Aral et~al.(2009)Aral, Muchnik, \protect\BIBand{}
  Sundararajan}]{aral2009distinguishing}
Aral S, Muchnik L, Sundararajan A (2009) Distinguishing influence-based
  contagion from homophily-driven diffusion in dynamic networks.
  \emph{Proceedings of the National Academy of Sciences of the United States of
  America} 106(51):21544--21549.

\bibitem[{Aral \protect\BIBand{} Walker(2012)}]{aral2012identifying}
Aral S, Walker D (2012) Identifying influential and susceptible members of
  social networks. \emph{Science} 337:337--341.

\bibitem[{Athey \protect\BIBand{} Imbens(2017)}]{athey2017econometrics}
Athey S, Imbens GW (2017) The econometrics of randomized experiments.
  \emph{Handbook of Economic Field Experiments}, volume~1, 73--140 (Elsevier).

\bibitem[{Backstrom et~al.(2006)Backstrom, Huttenlocher, Kleinberg,
  \protect\BIBand{} Lan}]{backstrom2006group}
Backstrom L, Huttenlocher D, Kleinberg J, Lan X (2006) Group formation in large
  social networks: membership, growth, and evolution. \emph{Proceedings of the
  12th ACM SIGKDD International Conference on Knowledge Discovery and Data
  Mining}, 44--54 (ACM).

\bibitem[{Bakshy et~al.(2012)Bakshy, Rosenn, Marlow, \protect\BIBand{}
  Adamic}]{bakshy2012role}
Bakshy E, Rosenn I, Marlow C, Adamic L (2012) The role of social networks in
  information diffusion. \emph{Proceedings of the 21st International Conference
  on World Wide Web}, 519--528 (ACM).

\bibitem[{Bloom et~al.(2015)Bloom, Liang, Roberts, \protect\BIBand{}
  Ying}]{bloom2015does}
Bloom N, Liang J, Roberts J, Ying ZJ (2015) Does working from home work?
  evidence from a chinese experiment. \emph{The Quarterly Journal of Economics}
  130(1):165--218.

\bibitem[{Bolton \protect\BIBand{} Ockenfels(2000)}]{bolton2000erc}
Bolton GE, Ockenfels A (2000) Erc: A theory of equity, reciprocity, and
  competition. \emph{American Economic Review} 90(1):166--193.

\bibitem[{Bolton \protect\BIBand{} Ockenfels(2006)}]{bolton2006inequality}
Bolton GE, Ockenfels A (2006) Inequality aversion, efficiency, and maximin
  preferences in simple distribution experiments: comment. \emph{American
  Economic Review} 96(5):1906--1911.

\bibitem[{Bond et~al.(2012)Bond, Fariss, Jones, Kramer, Marlow, Settle,
  \protect\BIBand{} Fowler}]{bond201261}
Bond RM, Fariss CJ, Jones JJ, Kramer AD, Marlow C, Settle JE, Fowler JH (2012)
  A 61-million-person experiment in social influence and political
  mobilization. \emph{Nature} 489(7415):295.

\bibitem[{Bott et~al.(2020)Bott, Cappelen, S{\o}rensen, \protect\BIBand{}
  Tungodden}]{bott2020you}
Bott KM, Cappelen AW, S{\o}rensen E{\O}, Tungodden B (2020) You’ve got mail:
  A randomized field experiment on tax evasion. \emph{Management Science}
  66(7):2801--2819.

\bibitem[{Brynjolfsson et~al.(2020)Brynjolfsson, Horton, Ozimek, Rock, Sharma,
  \protect\BIBand{} TuYe}]{brynjolfsson2020covid}
Brynjolfsson E, Horton JJ, Ozimek A, Rock D, Sharma G, TuYe HY (2020) Covid-19
  and remote work: An early look at us data. Technical report, National Bureau
  of Economic Research.

\bibitem[{Bulte et~al.(2018)Bulte, Wang, \protect\BIBand{}
  Zhang}]{bulte2018forced}
Bulte E, Wang R, Zhang X (2018) Forced gifts: The burden of being a friend.
  \emph{Journal of Economic Behavior \& Organization} 155:79--98.

\bibitem[{Cao et~al.(2020)Cao, Li, \protect\BIBand{} Liu}]{cao2020gift}
Cao C, Li SX, Liu TX (2020) A gift with thoughtfulness: A field experiment on
  work incentives. \emph{Games and Economic Behavior} 124:17--42.

\bibitem[{Centola \protect\BIBand{} Macy(2007)}]{centola2007complex}
Centola D, Macy M (2007) Complex contagions and the weakness of long ties.
  \emph{American Journal of Sociology} 113(3):702--734.

\bibitem[{Charness \protect\BIBand{} Rabin(2002)}]{charness2002understanding}
Charness G, Rabin M (2002) Understanding social preferences with simple tests.
  \emph{The Quarterly Journal of Economics} 117(3):817--869.

\bibitem[{Chen \protect\BIBand{} Li(2009)}]{chen2009group}
Chen Y, Li SX (2009) Group identity and social preferences. \emph{American
  Economic Review} 99(1):431--57.

\bibitem[{Christakis \protect\BIBand{} Fowler(2007)}]{christakis2007spread}
Christakis NA, Fowler JH (2007) The spread of obesity in a large social network
  over 32 years. \emph{New England Journal of Medicine} 357(4):370--379.

\bibitem[{Cialdini \protect\BIBand{} Goldstein(2004)}]{cialdini2004social}
Cialdini RB, Goldstein NJ (2004) Social influence: Compliance and conformity.
  \emph{Annual Review of Psychology} 55:591--621.

\bibitem[{Efron(1992)}]{efron1992bootstrap}
Efron B (1992) Bootstrap methods: Another look at the jackknife.
  \emph{Breakthroughs in Statistics}, 569--593 (Springer).

\bibitem[{Feldman(1984)}]{feldman1984development}
Feldman DC (1984) The development and enforcement of group norms. \emph{Academy
  of Management Review} 9(1):47--53.

\bibitem[{G{\"a}chter et~al.(2013)G{\"a}chter, Nosenzo, \protect\BIBand{}
  Sefton}]{gachter2013peer}
G{\"a}chter S, Nosenzo D, Sefton M (2013) Peer effects in pro-social behavior:
  Social norms or social preferences? \emph{Journal of the European Economic
  Association} 11(3):548--573.

\bibitem[{Holland \protect\BIBand{} Leinhardt(1971)}]{holland1971transitivity}
Holland PW, Leinhardt S (1971) Transitivity in structural models of small
  groups. \emph{Comparative Group Studies} 2(2):107--124.

\bibitem[{Holton(2001)}]{holton2001building}
Holton JA (2001) Building trust and collaboration in a virtual team. \emph{Team
  Performance Management: An International Journal} 7(3/4):36--47.

\bibitem[{Hossain \protect\BIBand{} Li(2014)}]{hossain2014crowding}
Hossain T, Li KK (2014) Crowding out in the labor market: A prosocial setting
  is necessary. \emph{Management Science} 60(5):1148--1160.

\bibitem[{Imai et~al.(2008)Imai, King, \protect\BIBand{}
  Stuart}]{imai2008misunderstandings}
Imai K, King G, Stuart EA (2008) Misunderstandings between experimentalists and
  observationalists about causal inference. \emph{Journal of the Royal
  Statistical Society: Series A} 171(2):481--502.

\bibitem[{Imbens \protect\BIBand{} Rubin(2015)}]{imbens2015causal}
Imbens GW, Rubin DB (2015) \emph{Causal Inference in Statistics, Social, and
  Biomedical Sciences} (Cambridge University Press).

\bibitem[{Jackson(2010)}]{jackson2010social}
Jackson MO (2010) \emph{Social and Economic Networks} (Princeton University
  Press).

\bibitem[{Kempe et~al.(2003)Kempe, Kleinberg, \protect\BIBand{}
  Tardos}]{kempe2003maximizing}
Kempe D, Kleinberg J, Tardos {\'E} (2003) Maximizing the spread of influence
  through a social network. \emph{Proceedings of the ninth ACM SIGKDD
  International Conference on Knowledge Discovery and Data Mining}, 137--146.

\bibitem[{Kernan et~al.(1999)Kernan, Viscoli, Makuch, Brass, \protect\BIBand{}
  Horwitz}]{kernan1999stratified}
Kernan WN, Viscoli CM, Makuch RW, Brass LM, Horwitz RI (1999) Stratified
  randomization for clinical trials. \emph{Journal of Clinical Epidemiology}
  52(1):19--26.

\bibitem[{Kizilcec et~al.(2018)Kizilcec, Bakshy, Eckles, \protect\BIBand{}
  Burke}]{kizilcec2018social}
Kizilcec RF, Bakshy E, Eckles D, Burke M (2018) Social influence and
  reciprocity in online gift giving. \emph{Proceedings of the 2018 CHI
  Conference on Human Factors in Computing Systems}, 126 (ACM).

\bibitem[{Liu et~al.(2015)Liu, Zhang, Chen, Guo, \protect\BIBand{}
  Yu}]{liu2015enterprise}
Liu S, Zhang Y, Chen L, Guo L, Yu D (2015) Enterprise wechat groups: Their
  effect on work-life conflict and life-work enhancement. \emph{Frontiers of
  Business Research in China} 9(4):516.

\bibitem[{Liu et~al.(2014)Liu, Yang, Adamic, \protect\BIBand{}
  Chen}]{liu2014crowdsourcing}
Liu TX, Yang J, Adamic LA, Chen Y (2014) Crowdsourcing with all-pay auctions: A
  field experiment on taskcn. \emph{Management Science} 60(8):2020--2037.

\bibitem[{Luo(2008)}]{luo2008changing}
Luo Y (2008) The changing chinese culture and business behavior: The
  perspective of intertwinement between guanxi and corruption.
  \emph{International Business Review} 17(2):188--193.

\bibitem[{Markovsky \protect\BIBand{} Lawler(1994)}]{markovsky1994new}
Markovsky B, Lawler EJ (1994) A new theory of group solidarity 11:113--137.

\bibitem[{Marsden(2002)}]{marsden2002egocentric}
Marsden PV (2002) Egocentric and sociocentric measures of network centrality.
  \emph{Social Networks} 24(4):407--422.

\bibitem[{Mauss(2002)}]{mauss2002gift}
Mauss M (2002) \emph{The gift: The Form and Reason for Exchange in Archaic
  Societies} (Routledge).

\bibitem[{McPherson et~al.(2001)McPherson, Smith-Lovin, \protect\BIBand{}
  Cook}]{mcpherson2001birds}
McPherson M, Smith-Lovin L, Cook JM (2001) Birds of a feather: Homophily in
  social networks. \emph{Annual Review of Sociology} 27(1):415--444.

\bibitem[{Newman et~al.(2006)Newman, Barabasi, \protect\BIBand{}
  Watts}]{newman2006structure}
Newman M, Barabasi AL, Watts DJ (2006) \emph{The Structure and Dynamics of
  Networks} (Princeton University Press).

\bibitem[{Nowak \protect\BIBand{} Roch(2007)}]{nowak2007upstream}
Nowak MA, Roch S (2007) Upstream reciprocity and the evolution of gratitude.
  \emph{Proceedings of the Royal Society B: Biological Sciences}
  274(1610):605--610.

\bibitem[{Park et~al.(2009)Park, Kee, \protect\BIBand{}
  Valenzuela}]{park2009being}
Park N, Kee KF, Valenzuela S (2009) Being immersed in social networking
  environment: Facebook groups, uses and gratifications, and social outcomes.
  \emph{Cyberpsychology \& behavior} 12(6):729--733.

\bibitem[{Pearl(2009)}]{pearl2009causality}
Pearl J (2009) \emph{Causality} (Cambridge University Press).

\bibitem[{Qiu et~al.(2016)Qiu, Li, Tang, Lu, Ye, Chen, Yang, \protect\BIBand{}
  Hopcroft}]{qiu2016lifecycle}
Qiu J, Li Y, Tang J, Lu Z, Ye H, Chen B, Yang Q, Hopcroft JE (2016) The
  lifecycle and cascade of wechat social messaging groups. \emph{Proceedings of
  the 25th International Conference on World Wide Web}, 311--320 (ACM).

\bibitem[{Roy(2005)}]{roy2005traditional}
Roy C (2005) \emph{Traditional festivals: a multicultural encyclopedia},
  volume~1 (Abc-Clio).

\bibitem[{Seinen \protect\BIBand{} Schram(2006)}]{seinen2006social}
Seinen I, Schram A (2006) Social status and group norms: Indirect reciprocity
  in a repeated helping experiment. \emph{European Economic Review}
  50(3):581--602.

\bibitem[{Shalizi \protect\BIBand{} Thomas(2011)}]{shalizi2011homophily}
Shalizi CR, Thomas AC (2011) Homophily and contagion are generically confounded
  in observational social network studies. \emph{Sociological Methods \&
  Research} 40(2):211--239.

\bibitem[{Siu(2001)}]{siu2001red}
Siu KWM (2001) Red packet: A traditional object in the modern world. \emph{The
  Journal of Popular Culture} 35(3):103--125.

\bibitem[{Stuart(2010)}]{stuart2010matching}
Stuart EA (2010) Matching methods for causal inference: A review and a look
  forward. \emph{Statistical Science} 25(1):1.

\bibitem[{Tajfel et~al.(1979)Tajfel, Turner, Austin, \protect\BIBand{}
  Worchel}]{tajfel1979integrative}
Tajfel H, Turner JC, Austin WG, Worchel S (1979) An integrative theory of
  intergroup conflict. \emph{Organizational Identity: A Reader} 56--65.

\bibitem[{Ugander et~al.(2012)Ugander, Backstrom, Marlow, \protect\BIBand{}
  Kleinberg}]{ugander2012structural}
Ugander J, Backstrom L, Marlow C, Kleinberg J (2012) Structural diversity in
  social contagion. \emph{Proceedings of the National Academy of Sciences}
  109(16):5962--5966.

\bibitem[{Veinott et~al.(1999)Veinott, Olson, Olson, \protect\BIBand{}
  Fu}]{veinott1999video}
Veinott ES, Olson J, Olson GM, Fu X (1999) Video helps remote work: Speakers
  who need to negotiate common ground benefit from seeing each other.
  \emph{Proceedings of the SIGCHI conference on Human Factors in Computing
  Systems}, 302--309.

\bibitem[{Wang et~al.(2008)Wang, Siu, \protect\BIBand{}
  Barnes}]{wang2008significance}
Wang CL, Siu NY, Barnes BR (2008) The significance of trust and renqing in the
  long-term orientation of chinese business-to-business relationships.
  \emph{Industrial Marketing Management} 37(7):819--824.

\bibitem[{Watts \protect\BIBand{} Strogatz(1998)}]{watts1998collective}
Watts DJ, Strogatz SH (1998) Collective dynamics of ‘small-world’ networks.
  \emph{Nature} 393(6684):440.

\bibitem[{Wheeler(1966)}]{wheeler1966toward}
Wheeler L (1966) Toward a theory of behavioral contagion. \emph{Psychological
  Review} 73(2):179.

\bibitem[{Yamagishi \protect\BIBand{} Cook(1993)}]{yamagishi1993generalized}
Yamagishi T, Cook KS (1993) Generalized exchange and social dilemmas.
  \emph{Social Psychology Quarterly} 235--248.

\bibitem[{Yuan et~al.(2021)Yuan, Altenburger, \protect\BIBand{}
  Kooti}]{yuan2021causal}
Yuan Y, Altenburger K, Kooti F (2021) Causal network motifs: Identifying
  heterogeneous spillover effects in {A}/{B} tests. \emph{Proceedings of the
  Web Conference 2021}, 3359--3370.

\end{thebibliography}





\end{document}